\documentclass[12pt]{article} 
 \usepackage{amsmath}
\usepackage{amssymb}
\usepackage{graphicx}
\usepackage{tikz}
\usepackage{xcolor}
\usepackage{framed}
\usepackage{bm}
  \usepackage{subfigure}
  \usepackage{caption}
 \usepackage{mciteplus}
 \usepackage{bbold}
   \definecolor{darkblue}{rgb}{0.1,0.1,.7}
 \usepackage[colorlinks, linkcolor=darkblue, citecolor=darkblue, urlcolor=darkblue, linktocpage,hyperfootnotes=false]{hyperref} 
 
 \numberwithin{equation}{section}
 
  \newlength{\abstractwidth}
  \newcommand{\be}{\begin{equation}}
  \newcommand{\bea}{\begin{eqnarray}}
  \newcommand{\eea}{\end{eqnarray}}
  \newcommand{\ea}{\end{eqnarray}}
  \newcommand{\beq}{\begin{equation}}
  \newcommand{\ee}{\end{equation}}
  \newcommand{\eeq}{\end{equation}}

\DeclareMathOperator{\Tr}{Tr}

\newcommand*{\lb}{\langle}
\newcommand*{\rb}{\rangle}

\newcommand*{\df}{\text{def}}

\newcommand{\mstr}{{\mathrm{MS}}}
\newcommand{\mm}{{\mathrm{MM}}}
\newcommand{\jt}{{\mathrm{JT}}}

  \newcommand{\aL}{2 \pi \big(1 - L (1 - \alpha)\big)}
  \newcommand{\ab}{\frac{\pi^2 \left(1 - L (1 - \alpha)\right)^2}{\beta}}

\begin{document}

\begin{titlepage}
 % \rightline{}
  \bigskip

  \bigskip\bigskip

  \bigskip

\begin{center}
%\centerline
{\Large \bf {Dilaton-gravity, deformations of the minimal string,}}\\
\vspace{0.5cm}
{\Large \bf {and matrix models}}
\bigskip

\end{center}

\bigskip
  \begin{center}

Gustavo J. Turiaci${}^1$, Mykhaylo Usatyuk${}^2$ and Wayne W. Weng${}^1$
  \bigskip \rm

{\small ${}^1$ Department of Physics, University of California, Santa Barbara, CA 93106, USA}  \\
\vspace{0.2cm}
{\small ${}^2$ Center for Theoretical Physics and Department of Physics, Berkeley, CA, 94720, USA}  \\
\rm

 % \bf {Write authors  }
  \bigskip \rm
\bigskip
 
\texttt{turiaci@ucsb.edu, musatyuk@berkeley.edu, wweng@ucsb.edu}
\rm

\bigskip
\bigskip

% \vspace{2cm}
  \end{center}

\vspace{4cm}
  \begin{abstract}
  
A large class of two-dimensional dilaton-gravity theories in asymptotically AdS$_2$ spacetimes are holographically dual to a matrix integral, interpreted as an ensemble average over Hamiltonians. Viewing these theories as Jackiw-Teitelboim gravity with a gas of defects, we extend this duality to a broader class of dilaton potentials compared to previous work by including conical defects with small deficit angles. In order to do this we show that these theories are equal to the large $p$ limit of a natural deformation of the $(2,p)$ minimal string theory.

 \medskip
  \noindent
  \end{abstract}
\bigskip \bigskip \bigskip

  \end{titlepage}

  %  \starttext \baselineskip=17.63pt \setcounter{footnote}{0}
   \tableofcontents

 % \sc

\newpage
\section{Introduction}

In recent years many insights regarding quantum gravity and black holes have been obtained by looking at simple models in two dimensions described by variants of Jackiw-Teitelboim (JT) dilaton-gravity \cite{Jackiw:1984je,*Teitelboim:1983ux,Almheiri:2014cka,Jensen:2016pah,*Maldacena:2016upp,*Engelsoy:2016xyb} that can be solved exactly \cite{Stanford:2017thb, Mertens:2017mtv,*Lam:2018pvp,Kitaev:2018wpr,Yang:2018gdb, Iliesiu:2019xuh, Saad:2019lba}. Of particular importance is the study of non-perturbative effects and spacetime wormholes, which connects pure JT gravity with a matrix integral \cite{Saad:2019lba} in the double-scaling limit \cite{Brezin:1990rb, *Douglas:1989ve, *Gross:1989vs}. This gives a new twist on holography where a bulk gravitational theory is related to an ensemble average over boundary Hamiltonians. This has been generalized in various directions, for example \cite{Maldacena:2019cbz, Cotler:2019nbi, Stanford:2019vob, Saad:2019pqd}.

It was argued in \cite{Saad:2019lba} that pure JT gravity, including non-perturbative effects, is equal to the large $p$ limit of the $(2,p)$ minimal string theory. This theory of 2D gravity has been known to be dual to a matrix integral for a long time \cite{Kazakov:1989bc, Staudacher:1989fy}. This correspondence has been further studied in \cite{Okuyama:2019xbv,*Okuyama:2020ncd,Johnson:2019eik,*Johnson:2020heh, Mertens:2020hbs, StanfordSeiberg, Mertens:2020pfe, Altland:2020ccq, Goel:2020yxl}. 

The connection between JT gravity and matrix integrals was generalized in \cite{Maxfield:2020ale, Witten:2020wvy} to include a gas of defects, which in turn can be related to more general 2D dilaton-gravity building on some previous work \cite{Mertens:2019tcm}). The goal of the present paper is to identify deformations of the minimal string that in a certain limit are equal to these deformations of JT gravity. Developing this connection will also allow us to find exact solutions for deformations of JT gravity that are outside the reach of the methods used by \cite{Maxfield:2020ale, Witten:2020wvy}. In the rest of this section we will give a brief summary of our results.

The $(2,p=2m-1)$ minimal string theory consists of coupling the $(2,p)$ 2D minimal model to 2D gravity. After fixing the conformal gauge this theory can be recast as a combination of the minimal model, the Liouville gravity mode $\phi$, and a set of $bc$ ghosts. We will study deformations of this theory described by the action 
\beq\label{intro:defactionms}
I= I_{(2,p)} - \sum_{n=1}^{m-1} \tau_n \int \mathcal{O}_{1,n} \hspace{0.1cm} e^{b(1+n)\phi},
\eeq
where $b=\sqrt{2/p}$ is the Liouville coupling. In this equation $ I_{(2,p)}$ represents the undeformed minimal model coupled to Liouville gravity. The constants $\tau_n$ are the couplings of the deformations labeled by a gravitationally dressed minimal model primary $\mathcal{O}_{1,n}$. The cosmological constant $\mu$ is identified with the $n=1$ deformation.

We will focus on the disk path integral with fixed length boundary conditions $\ell \equiv \oint_{\rm bdy} e^{b \phi}$. This can be computed using 2D CFT techniques, although in general this is difficult to do. Alternatively we can use the fact that the theory is dual to a matrix integral, with the matrix interpreted as a random Hamiltonian. All information is then encoded in the leading order disk density of states $\rho(E)$, where $E$ are the eigenvalues of the matrix (related to the boundary cosmological constant in the continuum description). Other observables and higher genus corrections are uniquely fixed by the loop equations \cite{Migdal:1975zf}. In a remarkable work, Belavin and Zamolodchikov \cite{Belavin:2008kv} proposed an exact expression for $\rho(E)$ valid to all orders in $\tau_n$. Their only input is the fact that the theory is equivalent to a matrix integral and that correlators on the sphere satisfy the fusion rules of the minimal model, following the program started in \cite{Moore:1991ir}.   

Another theory that is dual to a matrix integral is JT gravity with a gas of defects. These conical singularities are characterized by two numbers, a weighting factor $\lambda$ and a parameter $\alpha$ defined through the deficit angle $2\pi(1-\alpha)$. In general we have $0<\alpha<1$, with $\alpha=0$ being a cusp and $\alpha=1$ being basically no defect. This duality has been studied in \cite{Maxfield:2020ale, Witten:2020wvy} for the case $0<\alpha<1/2$ which we will refer to as sharp defects. The first result of this paper is to show, using the Belavin-Zamolodchikov solution, that the large $p$ limit of the deformed minimal string gives JT with a gas of defects. In this correspondence we identify each deformation with each defect species. The coupling $\tau_n$ is proportional to $\lambda$ in the large $p$ limit. We also scale the label of the minimal model operator as $n=\frac{p}{2}(1-\alpha)$, with fixed $\alpha$ identified as the other defect parameter. At finite $p$, $\alpha$ is a discrete parameter but becomes continuous at large $p$ and bounded between zero and one.

Even though we can check this connection between deformations of the minimal string and JT with a gas of defects by comparing explicit solutions of the theories, having a more direct argument would be preferable. In order to do this we can write the minimal string as a minimal model coupled to Liouville gravity. Then we can write a Lagrangian representation of the minimal model as time-like Liouville (this connection is not completely understood; see \cite{KapecMahajan} for a recent discussion) and a field redefinition gives JT with defects (this is a simple generalization of \cite{StanfordSeiberg, Mertens:2020hbs}). We summarize the relation between these theories in figure \ref{fig:triangle}.

Taking the large $p$ limit of the minimal string solution, we find the exact disk density of states for JT gravity coupled to general defects. The answer we obtain from the Belavin-Zamolodchikov solution is given in \eqref{JTdefgeneralsemul}. Instead, we will present the result using a trick pointed out to us by T. Budd \cite{Budd}. It is convenient to define a defect generating function $W(y)$ as
\beq\label{intro:dgf}
W(y) \equiv \sum_i \lambda_i e^{-2\pi (1-\alpha_i)y},
\eeq
where $i$ is an index labeling defect species, which can be continuous. Providing a function $W(y)$ is equivalent to specifying the angle and weights of the defects. Since $0<\alpha_i<1$ the inverse Laplace transform of $W(y)$ should have support on $(-2\pi,0)$. 

Before presenting the solution to the disk density of states we need to specify the edge of the spectrum $E_0$ where $\rho(E<E_0)=0$. In terms of the defect generating function, the large $p$ limit of the Belavin-Zamolodchikov solution \eqref{JTdefgeneralsemul} gives $E_0$ as the largest solution of
\beq
 \int_{\mathcal{C}} \frac{dy}{2\pi i} e^{2\pi y} \left( y-\sqrt{y^2 - 2 W(y) - E_0 }\right)=0.
 \eeq
 Using this result, the disk density of states for $E>E_0$, obtained in the same way, is given by
 \beq 
 \rho(E) = \frac{e^{S_0}}{2\pi} \int_{\mathcal{C}} \frac{dy}{2\pi i} e^{2\pi y} \tanh^{-1}\left(\sqrt{\frac{E-E_0}{y^2-2W(y)-E_0} }\right).
\eeq
The contour $\mathcal{C}$ is the one appropriate for an inverse Laplace transform, running along the imaginary direction with a real part such that all singularities are to the left. This solution matches the one found in \cite{Maxfield:2020ale, Witten:2020wvy} when $W$ involves only defects in the range $0<\alpha\leq1/2$. Since the connection between the minimal string deformations, defects, and Belavin-Zamolodchikov solution are valid for any value of $\alpha$ we claim that this solution is valid for JT gravity with a gas of general defects with $0<\alpha<1$. The solution we find for $\alpha>1/2$ is very different from the $0<\alpha\leq1/2$ solution, analytically continued in $\alpha$. This feature is most transparent in \eqref{JTdefgeneralsemul}. The new terms we find have a nice geometrical interpretation as we explain later on\footnote{When $\alpha>1/2$ there is the possibility of defects merging and this produces new contributions to the density of states. This is reminiscent of the situation with conical defects in 3D gravity and 2D CFT associated to operators with $h<(c-1)/32$ \cite{Collier:2018exn}. We thank S. Collier for discussions on this.}. It is an open problem to derive this result using the JT gravity path integral representation of the theory, since the methods of \cite{Maxfield:2020ale, Witten:2020wvy} cannot be directly applied for reasons we review in next section.
\begin{figure}
\begin{center}
\begin{tikzpicture}[scale=0.5]
\draw[thick] (-4,0) -- (4,0) -- (0,6.9282) -- (-4,0);
\draw[blue] (-7,-1) node {\small $(2,p)$ minimal string};
\draw[blue] (-7,-1.9) node {\small $+$ deformations};
\draw[blue] (7,-1) node {\small time-like $+$ space-like};
\draw[blue] (7,-1.9) node {\small Liouville};
\draw[blue] (0,7.9282) node {\small JT $+$ defects};
\draw (-5.6,3.65) node {\small Compare $\rho_{\rm disk}(E)$};
\draw (5.6,3.65) node {\small Field redefinition};
\draw (0,-1) node {\small Coulomb gas};
\end{tikzpicture}\hspace{1.2cm}
\end{center}
\vspace{-0.5cm}
\caption{Relation between the three theories we are interested in and their respective deformations: the minimal string, time-like Liouville coupled to space-like Liouville and JT gravity with defects. The connection between the $(2,p)$ minimal string and the combination of Liouville theories is the least rigorous since it relies on a Coulomb gas representation of the minimal model.}\label{fig:triangle}
\end{figure}

Finally, we study the connection between JT gravity with defects and the quantization of 2D dilaton-gravity
\beq
I= -\frac{1}{2} \int \sqrt{g} \left[ \Phi R + 2 U(\Phi) \right].
\eeq
In the simplest quantization scheme, as explained in \cite{Witten:2020wvy}, one can identify the dilaton potential with the defect generating function as $U(\Phi) = \Phi + W(\Phi)$. This can be justified by Taylor expanding the path integral in powers of $W$ representing contributions from an arbitrary number of defects. The minimal string gives a new perspective on this connection. In order to do this we start from the action \eqref{intro:defactionms} and rewrite the minimal model in a Coulomb gas, or time-like Liouville, representation. This action can be combined with the gravitational Liouville mode into a 2D metric and a 2D dilaton which we identify with $\Phi$ \cite{StanfordSeiberg,Mertens:2020hbs}. This approach to the quantization scheme gives a 2D dilaton-gravity theory with a slightly different identification between the parameters $\lambda, \alpha$ and the dilaton potential, see equation \eqref{eq:minstrindilpot}. We use this approach to propose a solution of dilaton-gravity with a polynomial potential. The advantage of this scheme is that it gives an answer consistent with a semiclassical limit. Some recent studies on other aspects of dilaton-gravity are \cite{Anninos:2020cwo, Momeni:2020zkx, *Momeni:2020tyt, MeffordSuzuki, Narayan:2020pyj,Alishahiha:2020jko}.

The organization of the rest of the paper is as follows. In section \ref{sec:JTdefreview} we review the solution for JT gravity with a gas of sharp defects and its connection to 2D dilaton-gravity. In section \ref{sec:minimalstring} we describe the minimal string theory and its deformations, and present the exact solution proposed by Belavin-Zamolodchikov. In section \ref{sec:JTsection} we show the connection between these two theories. We use the minimal string to find a solution of JT gravity with a gas of defect with arbitrary angles. In section \ref{sec:dilgrav} we explain the connection with dilaton-gravity. We finish presenting conclusions and open directions in section \ref{sec:discussions}. We leave technical details for Appendices.

\textbf{Note:} While this work was in progress, we became aware of a related work by T. Budd \cite{Budd}. The author found the same results by computing the Weil-Petersson volumes via a geometric construction. 

\section{JT gravity with defects: Review} \label{sec:JTdefreview}
We are interested in analyzing the path integral of JT gravity with conical defects. We first review the genus expansion of pure JT gravity following \cite{Saad:2019lba}, followed by a summary of JT gravity with a gas of ``sharp" defects with deficit angles $\pi < \theta\equiv 2\pi (1-\alpha) < 2\pi$, or equivalently $0<\alpha<1/2$  \cite{Maxfield:2020ale,Witten:2020wvy} \footnote{We follow the conventions in \cite{Maxfield:2020ale},  related to \cite{Witten:2020wvy} via $\alpha_{\rm W} \to 1-\frac{\alpha_{\rm MT}}{2\pi}$.}. These theories were shown to be dual to certain double-scaled matrix models. We will comment on this briefly but postpone the full discussion to sections \ref{sec:minimalstring} and \ref{sec:JTsection}, when we discuss defects with general deficit angles $0 < \theta < 2\pi$, or equivalently $0<\alpha<1$, and the connection to the deformed minimal string. We will refer to the range $\alpha>1/2$ as ``blunt" defects. The case $\alpha=1/2$ is treated separately in section \ref{sec:JTsection}, where we show that it exhibits the same behavior as $\alpha < 1/2$.

\subsection{Pure JT gravity}
JT gravity in Euclidean signature is a 2D dilaton-gravity theory described by a metric $g$ coupled to a scalar dilaton $\Phi$ with action
\be
I_\jt = -\frac{S_0}{2\pi} \left( \frac{1}{2} \int_M \sqrt{g} R + \int_{\partial M} \sqrt{h} K \right) -\frac{1}{2} \int_M \sqrt{g} \Phi (R+2) - \int_{\partial M} \sqrt{h} \Phi (K-1),
\ee
where we have set $\ell_{AdS} = 1$. The first term is the Einstein-Hilbert term, which by the Gauss-Bonnet theorem is equivalent to the Euler characteristic $\chi(M)$ of the manifold, so it is purely topological. For a surface with genus $g$ and $n$ boundaries, $\chi = 2-2g-n$. We will assume $S_0$ is some large parameter\footnote{When the theory is regarded as an effective action for the near-horizon dynamics of near-extremal black holes, the prefactor $S_0$ has the interpretation as the Bekenstein-Hawking entropy.} so that $e^{-S_0}$ plays the role of suppressing higher genus topologies in the path integral and the theory can be studied under an asymptotic ``genus expansion." The dilaton $\Phi$ appears linearly in the action and merely acts as a Lagrange multiplier enforcing the constraint $R+2=0$. This fixes the bulk manifold $M$ to be a patch of hyperbolic surface, bounded by some boundary curve $\partial M$.

Thus, all of the non-trivial dynamics come from the boundary term. For geometries with circular asymptotic boundaries we choose boundary conditions such that the proper length of each boundary is fixed to be $\beta/\epsilon$ where $\epsilon$ is regarded as a ``holographic" regulator which is eventually taken to zero. The dilaton is fixed to be a constant $\Phi_b = \gamma/\epsilon$ on each boundary. Taking $\epsilon \rightarrow 0$ while keeping the ratio $\beta/\gamma$ fixed corresponds to sending $\partial M$ to the asymptotic boundary. We set $\gamma=1/2$ for simplicity. This gives the familiar Schwarzian action \cite{Jensen:2016pah}, which encodes fluctuations of the curve.
\begin{figure}
    \centering
    \includegraphics[scale=0.5]{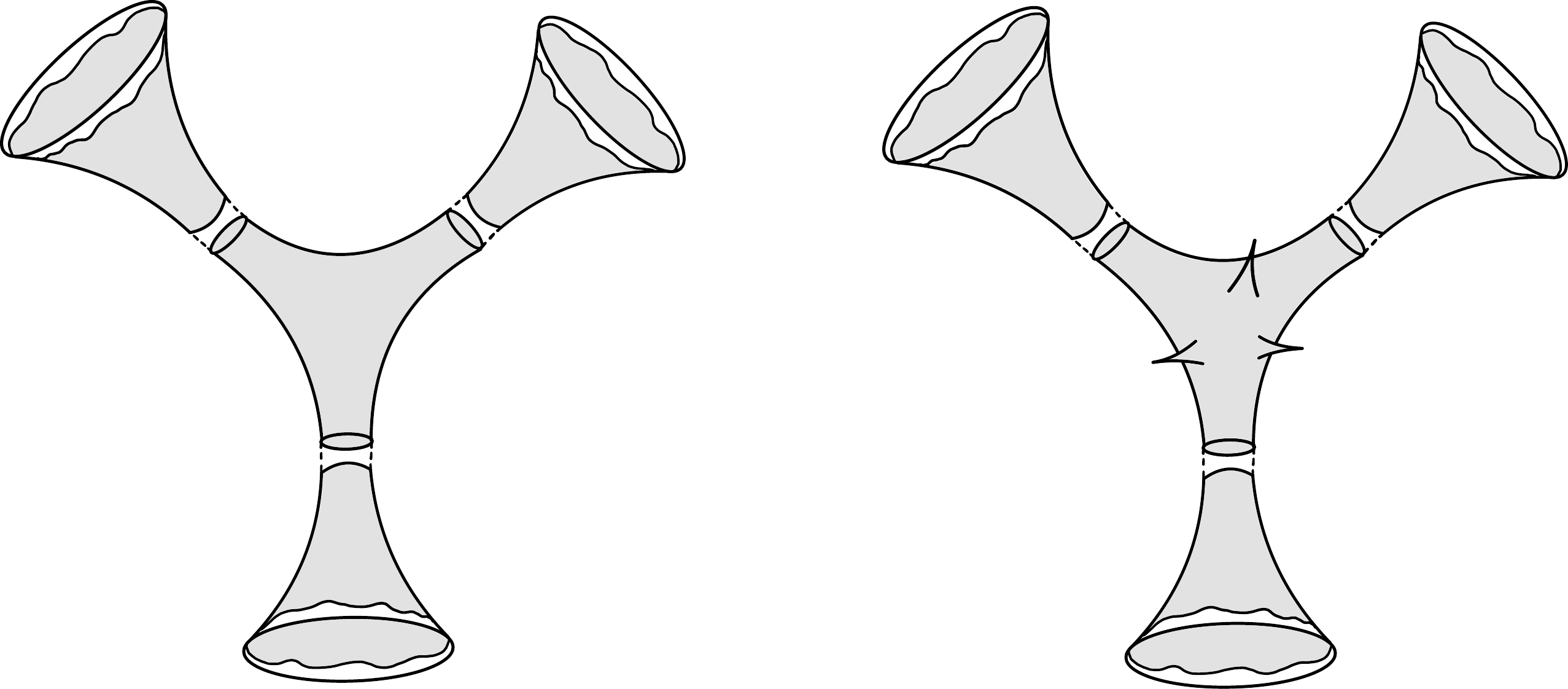}
    \caption{\textbf{Left:} The SSS construction of the correlator $\langle Z(\beta_1) Z(\beta_2) Z(\beta_3) \rangle_C$. The Weil-Petersson volume $V_{g=0,n=3}$ is glued to trumpets along each of the three geodesic boundaries. \textbf{Right:} Analogous construction of the correlator with sharp conical defects. The Weil-Petersson volume $V_{g=0,n=3,k=3}$ has three sharp defects.}
    \label{fig:sss-defect}
\end{figure}

We are interested in path integrals over connected geometries with $n$ asymptotic boundaries of regularized lengths $\beta_i$. We will denote this quantity by the $n$-point connected correlator $\lb Z(\beta_1)...Z(\beta_n)\rb_C$. The topological term in the action naturally organizes the path integral into a topological expansion in the genus $g$ of bulk manifolds. A manifold of genus $g$ with $n$ asymptotic boundaries has Euler characteristic $\chi = 2-2g-n$ so the genus expansion has the form
\be
\lb Z(\beta_1)...Z(\beta_n)\rb_C = \sum_{g=0}^\infty e^{-S_0 (2g+n-2)} Z_{g,n}(\beta_1,...,\beta_n).
\ee
In \cite{Saad:2019lba} an explicit expression for $Z_{g,n}$ was found in terms of volumes of moduli spaces of hyperbolic Riemann surfaces with geodesic boundaries. It will turn out that a similar construction generalizes to JT with sharp conical defects, so it will be useful to first understand the pure JT construction.

To compute $Z_{g,n}$ we must integrate over all constant negative curvature geometries of genus $g$ with $n$ asymptotic boundaries. All of these geometries can be constructed by gluing $n$ trumpet geometries to an internal surface of genus $g$ with $n$ geodesic boundaries. See figure \ref{fig:sss-defect}. We can now fix the lengths of the geodesic boundaries to be $b_1,...,b_n$ and compute the contributions of the trumpets and internal geometry separately. Finally, we can integrate these contributions for all geodesic boundary lengths, with appropriate measure, to recover $Z_{g,n}$ as desired. 

The path integral on each trumpet is
\be
Z_{\text{trumpet}}(\beta,b) = \sqrt{\frac{1}{4\pi \beta}} e^{-\frac{ b^2}{4 \beta}}. 
\ee
We must also integrate over all internal geometries of genus $g$ with $n$ geodesic boundaries of lengths $b_i$. These surfaces form the moduli space $\mathcal{M}_{g,n}(b_1,...,b_n)$ of bordered Riemann surfaces. The proper measure for this moduli space is provided by the Weil-Petersson (WP) measure, and the integral over the moduli space reduces to the Weil-Petersson volume $V_{g,n}(b_1,...,b_n)$ \cite{Mirzakhani:2006fta}. Putting everything together, we integrate over all possible geodesic lengths $b$ with the proper gluing measure $b db$ to get all stitchings of trumpets with internal moduli with the final result
\be
Z_{g, n}(\beta_1, ..., \beta_n)=\int_0^\infty b_1 d b_1 Z_{\text {trumpet}}(\beta_1, b_1) \cdots \int_0^\infty b_{n} d b_{n} Z_{\text {trumpet}}(\beta_n, b_n) V_{g, n}(b_1, ..., b_n).
\ee

There are two special cases for which the above formula does not apply, the disk $Z_{0,1}(\beta)$ and the double trumpet $Z_{0,2}(\beta_1,\beta_2)$. It will be interesting to compare these amplitudes to the corresponding ones with defects so we quote the pure JT results here:
\be
Z_{0,1}(\beta)=\sqrt{\frac{1}{16 \pi \beta^{3}}} e^{ \frac{\pi^2}{\beta}}, \quad Z_{0,2}(\beta_1,\beta_2) = \frac{\sqrt{\beta_1\beta_2}}{2\pi (\beta_1+\beta_2)}.
\ee
The structure of the genus expansion implies that JT gravity is dual to a double-scaled Hermitian matrix model \cite{Saad:2019lba}. An integral transform of the JT amplitudes $Z_{g,n}$ satisfies the topological recursion property of matrix integrals, derived from the loop equations. This turns out to be a consequence of Mirzakhani's recursion relations \cite{Mirzakhani:2006fta} for $V_{g,n}$, which were shown to be equivalent to a topological recursion by Eynard and Orantin \cite{Eynard:2007fi}. The gravitational path integral can now be interpreted as a matrix integral
\be
\lb Z(\beta_1)...Z(\beta_n) \rb = \int d H e^{-L \Tr V(H)} \Tr(e^{-\beta_1 H}) ... \Tr(e^{-\beta_n H}),
\ee
where $H$ is a Hermitian matrix of size $L \times L$. Formally we take a double-scaling limit on the right-hand side where we send $L \rightarrow \infty$ while tuning the potential $V(H)$ such that the matrix model density of states matches the JT density of states at leading order. The matrix $H$ is interpreted as the Hamiltonian of a dual quantum system and we interpret JT gravity as an ensemble average over independent quantum systems.

\subsection{Sharp defects} \label{sec:sharpdef1}
We will now briefly review the work of \cite{Maxfield:2020ale,Witten:2020wvy} on JT gravity with conical defects of deficit angle $0 < \alpha < 1/2$. The reason for this restriction will be apparent shortly.

We are again interested in path integrals over connected geometries with $n$ asymptotic boundaries which we denote by $\lb Z(\beta_1)...Z(\beta_n)\rb_C$, but allowing for the presence of a gas of defects. The path integral naturally organizes into a topological expansion in genus alongside an expansion in the number of defects inserted in the bulk
\be \label{eqn:JTDefectParitionfn}
\lb Z(\beta_1)...Z(\beta_n)\rb_C = \sum_{g=0}^\infty \sum_{k=0}^{\infty} e^{-S_0 (2g+n-2)} \frac{\lambda^k}{k!} Z_{g,n,k}(\beta_1,...,\beta_n).
\ee
The term proportional to $\lambda^k$ inserts $k$ conical defects in the bulk integrated over all possible insertion positions. The defects are indistinguishable so the symmetry factor $k!$ prevents overcounting identical configurations\footnote{The moduli space and volumes $V_{g,n,k}$ are typically defined with distinguishable points, so $Z_{g,n,k}$ is defined here with distinguishable defects.}. From this expression we see the weight $\lambda$ acts as a fugacity with respect to the number of defects. 

The factor $Z_{g,n,k}$ consists of an integral over all constant negative curvature geometries of genus $g$ with $k$ conical points and $n$ asymptotic boundaries. It turns out that for conical defects with deficit angles $\pi < \theta < 2\pi$, or equivalently $0< \alpha < 1/2$, all such geometries can be constructed by gluing trumpets to internal geometries of genus $g$ with $k$ conical points \cite{tan2004generalizations}, the only exception being the disk with one defect, where a direct calculation \cite{Mertens:2019tcm} gives $Z_{0,1,k}= Z_{\rm trumpet}(b=2\pi i \alpha)$. Besides this exception, all of our formulas from the pure JT discussion carry over as long as we replace the moduli space that we are integrating over in
\begin{align}
& Z_{g, n, k}\left(\beta_{1},\dots, \beta_{n}; \alpha_{1},\dots, \alpha_{k}\right) = \\
    & \quad \int_{0}^{\infty} b_{1} d b_{1} \, Z_{\text {trumpet}}\left(\beta_{1}, b_{1}\right) \dots \int_{0}^{\infty} b_{n} d b_{n} \,Z_{\text {trumpet}}\left(\beta_{n}, b_{n}\right) V_{g, n, k}\left(b_{1},\dots, b_{n}; \alpha_{1}, \dots, \alpha_{k}\right), \nonumber
\end{align}
with $V_{g,n,k}$ the Weil-Petersson volumes of the moduli space of surfaces of genus $g$ with $n$ geodesic boundaries and $k$ conical points. For $\alpha < 1/2$ there is a simple relation between the WP volumes with conical points $V_{g,n,k}$ and volumes without conical points $V_{g,n+k}$ \cite{tan2004generalizations,do2006weilpetersson,do2011moduli}. The WP volumes with $k$ conical points can be found from the ordinary volumes by analytically continuing $k$ of the $n+k$ geodesic boundary lengths to imaginary values
\be
V_{g, n, k}\left(b_{1}, \ldots b_{n} ; \alpha_{1}, \ldots, \alpha_{k}\right)=V_{g, n+k}\left(b_{1}, \ldots b_{n}, b_{n+1}=2 \pi i \alpha_{1}, \ldots, b_{n+k}=2 \pi i \alpha_{k}\right).
\ee
Using the above relation, and a formula for the genus zero WP volumes previously derived in \cite{Mertens:2020hbs}, reference \cite{Maxfield:2020ale} re-summed the defect expansion at genus zero
\be
\lb Z(\beta_1)...Z(\beta_n)\rb_{C,g=0} = e^{-S_0 (n-2)} \sum_{k=0}^{\infty} \frac{\lambda^k}{k!} Z_{0,n,k}(\beta_1,...,\beta_n).
\ee
We will give the answer for the case with multiplet species of sharp defects $(\lambda_i, \alpha_i)$. Performing the sum explicitly and doing an inverse Laplace transform gives the density of states 
\be
\langle\rho(E)\rangle_{g=0}=\frac{ e^{S_{0}}}{2 \pi} \int_{ E_{0}}^{E} \frac{d u}{\sqrt{ E-u}}\left(I_{0}(2 \pi \sqrt{u})+\sum_{i} \lambda_{i} \frac{2 \pi \alpha_{i}}{\sqrt{u}} I_{1}\left(2 \pi \alpha_{i} \sqrt{u}\right)\right),
\ee
where $I_n$ are modified Bessel functions of the first kind. Here, $E_0(\lambda)$ gives the spectral edge of the distribution and is the largest root of the string equation which we review in section \ref{subsection:stringeqn}. We will derive the same result from the minimal string in section \ref{sec:JTsection}.

JT gravity with defects is again dual to a matrix integral which can be shown by proving that integral transforms of $Z_{g,n,k}$ satisfy topological recursion. We omit the details, arriving at the same result through the minimal string in the next section.

We finish by outlining the connection to dilaton-gravity. It was argued in \cite{Mertens:2019tcm} that an insertion of a defect with parameters $(\lambda,\alpha)$ in JT gravity is equivalent to the insertion in the path integral of  
\be
\lambda \int d^2x \sqrt{g} e^{-2\pi (1-\alpha)\Phi}.
\ee
By summing over an arbitrary number of defects and summing over possible defect species this insertion simply exponentiates \cite{Maxfield:2020ale,Witten:2020wvy}. The end result is a modification of the action of JT gravity into a more general dilaton-gravity model
\be
I = -\frac{1}{2}\int d^2x \sqrt{g} \left( \Phi R + 2 U(\Phi)\right), ~~~ U(\Phi) = \Phi - \sum_i \lambda_i \, e^{-2\pi(1-\alpha_i)\Phi},
\ee
where the sum $i$ is over the defect species. As emphasized in \cite{Witten:2020wvy} there is a choice of renormalization scheme behind this identification between a quantum theory (the sum over defects) and a classical dilaton potential. We will give a different scheme using the minimal string in section \ref{sec:dilgrav}.

\subsection{General defects} \label{sec:generaldefects}

For blunt defects, corresponding to $\alpha>1/2$, the previous methods cannot be used for the reasons we will emphasize now. The recipe of \cite{Saad:2019lba} involves first integrating out the dilaton and summing over hyperbolic surfaces with cone points. The second step involves finding geodesics homologous to the holographic boundaries and computing the path integral by gluing Weil-Petersson volumes with `trumpet' partition functions. This step fails when $\alpha>1/2$ since hyperbolic surfaces with these cone points do not necessarily always have geodesics that we could use to cut and glue. 

\begin{figure}
    \centering
    \includegraphics[scale=0.5]{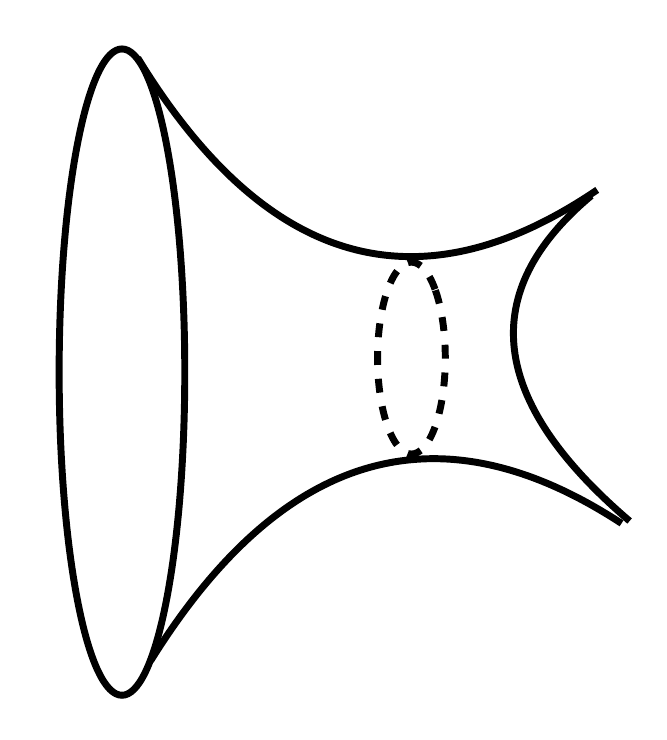}\hspace{4cm}
        \includegraphics[scale=0.46]{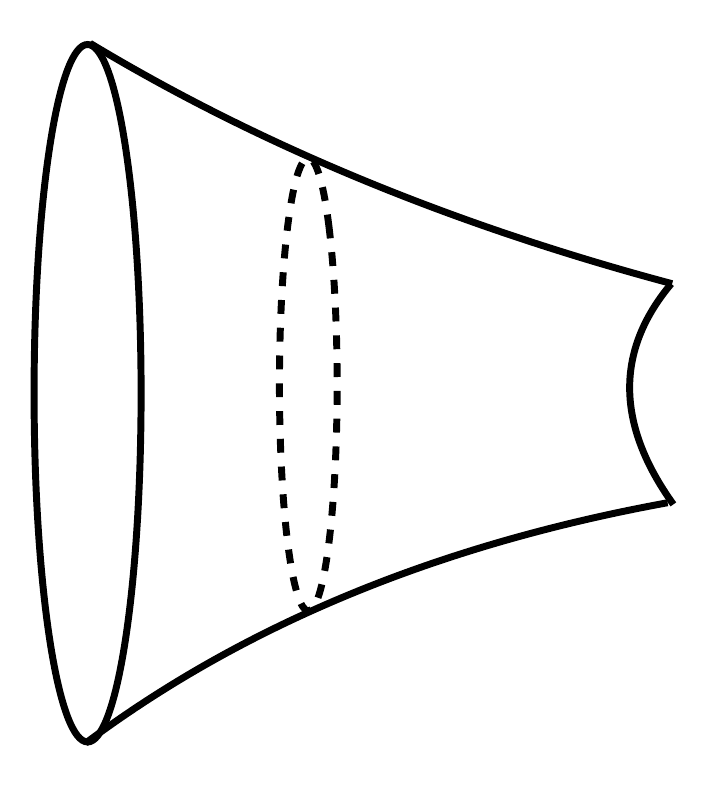}
    \caption{\textbf{Left:} Picture of a hyperbolic surface with a holographic boundary to the left and two sharp defects shown to the right with $\alpha<1/2$. In dashed line we show a geodesic homologous to the holographic boundary. \textbf{Right:} Similar picture of a hyperbolic surface in the case $\alpha>1/2$. As argued in the main text there is no geodesic to cut and glue.}
    \label{fig:geonogeo}
\end{figure}

There is a simple argument explaining why this is the case. Assume the opposite, namely that we can write a hyperbolic surface with $k$ cone points labeled by $\alpha_i$ and for concreteness take a single geodesic boundary and no handles. We can use the Gauss-Bonnet formula on a two-dimensional surface with $k$ cone points and with geodesic boundaries
\beq
\frac{1}{2} \int_M R + \sum_{i=1}^k 2\pi(1-\alpha_i) = 2\pi \chi(M).
\eeq
For surfaces with $R=-2$ away from cone points and one geodesic boundaries this gives the inequality $\sum_{i=1}^k (1-\alpha_i)> 1$. When $\alpha_i<1/2$, this inequality is always satisfied as long as we have more than a single defect. Instead, assume we have a single defect species with $\alpha>1/2$. Then we need at least $k > 1/(1-\alpha)$ in order for the inequality to be satisfied. We show this in Figure \ref{fig:geonogeo}. 

Another manifestation of the fact that general defects are special is the fact that they can merge without having to pinch off the surface. Consider the case of a single species of defects with defect parameter $\alpha$. In general, $k$ of them can merge into a single ``large" defect with cone angle
\beq
2\pi\alpha^{(k)} = 2\pi(1-k(1-\alpha))
\eeq
as long as $\alpha^{(k)}$ is positive. This is not possible when $\alpha < 1/2$, so sharp defects do not merge in a smooth way. Instead, when $\alpha > 1/2$ we can merge no more than $k = \lfloor \frac{1}{1-\alpha} \rfloor$ defects. 

For these reasons instead of following the approach of \cite{Saad:2019lba}, we will exploit the connection between JT gravity with defects and the minimal string that we develop in section \ref{sec:minimalstring}. The solution of the deformed minimal string given in \cite{Belavin:2008kv} is insensitive to whether it corresponds to sharp defects or not. Therefore we will find the JT answer by taking a limit of the minimal string.   

\section{Deformations of the minimal string} \label{sec:minimalstring}
In this section we review the relevant aspects of the $(2,p)$ minimal string and its deformations by adding ``tachyon" operators in the action. These string theories are dual to a Hermitian one-matrix integral in the double-scaling limit. Therefore the tree-level string equation of the matrix integral, or equivalently the disk density of states, completely specifies the model in the double-scaling limit. The most general tree-level string equation corresponding to the deformed minimal string was proposed by Belavin and Zamolodchikov \cite{Belavin:2008kv} building on the work \cite{Moore:1991ir}.

\subsection{The minimal string} \label{sec:minstringasdilatongrav}
We will begin with a brief review of the minimal string theory. We will interpret it as a theory of 2D gravity on the worldsheet coupled to a minimal model CFT. In the conformal gauge the physical metric $g$ can be written in terms of a fiducial metric $\hat{g}$ and a scale factor as $g= e^{2 b \phi} \hat{g}$, with $b$ a parameter to be determined later. After gauge fixing, the minimal string reduces to a minimal model coupled to Liouville field theory associated to the scale factor $\phi$, and $bc$ ghosts. Below we review some useful facts of these building blocks that we will need later.

\paragraph{Minimal model:} The minimal string can be defined for any minimal model labeled by coprime integers $(p,p')$. Since we will be interested in the theories dual to a one-matrix integral we focus on the Lee-Yang series $(2,p=2m-1)$, for integer $m$. The central charge of these theories is given by $c_M=1-6 q^2$, where we define $q=1/b-b$ and $b=\sqrt{2/p}$. The interpretation of these parameters will become clear soon. We will consider theories with odd $m$ since only those exist non-perturbatively; see for example \cite{Maldacena:2004sn}.  

The spectrum of these models includes a finite number of primaries labeled by a single integer with scaling dimension
\beq
\mathcal{O}_{1,n},~~~~{\rm with}~~\Delta_n = \frac{(n b^{-1} -b)^2-(b^{-1}-b)^2}{4},~~~~n=1,\ldots, m-1.
\eeq
These primary operators satisfy very simple fusion rules. Using the Coulomb gas approach, some aspects of the minimal models can be reproduced by a time-like Liouville action, as discussed by Zamolodchikov \cite{Zamolodchikov:2005fy} (see also \cite{Kostov:2005kk}), in terms of a scalar field $\chi$. The action is given by 
\beq \label{eqn:timelikeaction}
I_{\mm}[\chi]= \frac{1}{4\pi} \int  \sqrt{\hat{g}} \left[-(\hat{\nabla} \chi)^2 - q \hat{R} \chi - 4\pi \mu e^{2 b \chi} \right],
\eeq
where $\mu$ is a parameter we will fix later. This gives a Lagrangian description of these CFT. This representation motivates the connection to dilaton-gravity following \cite{StanfordSeiberg}. Within this time-like Liouville theory consider the following operators
\beq
\exp{\big(2 \hat{a}_n \chi \big)},~~~~~\hat{a}_n =\frac{b}{2}(1-n),~~~~~n=1,\ldots,m-1,
\eeq
with scaling dimension $\Delta_{\hat{a}_n}=\hat{a}_n(q+\hat{a}_n)$. We associate these operators to the minimal model primaries $\mathcal{O}_{1,n} \leftrightarrow \exp{\big(2 \hat{a}_n \chi \big)}$, although with a different normalization we present below. It was verified in \cite{Zamolodchikov:2005fy} that some correlators of time-like Liouville reproduce the minimal model correlators, after imposing the fusion rules by hand. Some recent progress understanding this identification can be found in \cite{KapecMahajan} (for a different approach see \cite{Kostov:2002uq,*Kostov:2003uh}). 
 
 \paragraph{Liouville sector:} The second building block of the minimal string is the Liouville gravity mode. This is an action for the scale factor $\phi$ in the conformal gauge, originating from the conformal anomaly, given by
\beq
I_L[\phi]= \frac{1}{4\pi} \int d^2 z \sqrt{\hat{g}} \left[(\hat{\nabla} \phi)^2 + Q \hat{R} \phi + 4\pi \mu e^{2 b \phi} \right],
\eeq
where $Q=b+1/b$. The central charge is given by $c_L=1+6Q^2$. The cancellation of the conformal anomaly between the minimal model, Liouville and the ghosts fixes the parameter $b=\sqrt{2/p}$. The primary operators are
\beq
V_a = \exp{\big( 2 a \phi \big)},~~~~~~a=\frac{Q}{2}+i P,~~~~~\Delta_a = a(Q-a)
\eeq
which we can write in terms of $a$ or the Liouville momentum $P$, which is continuous. The normalizable states correspond to $P^2>0$ while non-normalizable local operator insertions have $P^2<0$. We will be interested in the case of geometries with boundaries. In this case we will define the FZZT boundary condition \cite{Fateev:2000ik} with fixed boundary cosmological constant 
\beq
\mu_B = \kappa \cosh\left(2 \pi b s \right) ,~~~\kappa \equiv \frac{\sqrt{\mu}}{\sqrt{\sin \pi b^2}},
\eeq
in terms of the parameter $s$. Equivalently we will work with the fixed length boundary condition which is a Laplace transform of the FZZT brane. This can be thought of a Dirichlet boundary condition on the Liouville field that fixes the value of $\oint e^{b\phi} \to \ell$ along the boundary. For a detailed explanation of how to go between them see \cite{Mertens:2020hbs}.

\paragraph{Minimal string:} The minimal string theory is a combination of the minimal model CFT, the Liouville mode and a set of ghosts. These sectors are only coupled through the anomaly cancellation, and a possible integral over moduli space when computing observables. We will only consider in this paper states of ghost number one. These are ``tachyon" operators that are given by gravitationally dressing minimal model primaries
\beq\label{skrkw}
\mathcal{T}_n \equiv \int \sqrt{\hat{g}} \hspace{0.1cm} \mathcal{O}_{1,n} \hspace{0.07cm} e^{2 a_n \phi},~~~~~a_n = \frac{b}{2}(1+n),~~~~n=1,\ldots,m-1.
\eeq
For the operator $\mathcal{T}_{n}$ to be diffeomorphism invariant, the composite operator $\mathcal{O}_{1,n} \hspace{0.07cm} e^{2 a_n \phi}$ 
should have dimension $(1,1)$. The dressing parameter was determined from the condition
\beq
\Delta_{n} + \Delta_{a_n}=1.
\eeq
There is another solution for $a_n$ to this equation, but in \eqref{skrkw} we picked the one that is smooth in the $b\to0$ limit. 

In the rest of this section we will review a connection pointed out by Seiberg and Stanford \cite{StanfordSeiberg} (see also Appendix F of \cite{Mertens:2020hbs}) between this theory and two-dimensional dilaton-gravity that we will exploit later on. The idea is to use the time-like Liouville representation of the minimal model. Then the action of the minimal string can be written as a sum of two Liouville modes $I_{\mstr}[\chi,\phi]=I_{\mm}[\chi] + I_\text{L}[\phi]+I_{\text{ghosts}}$. The ghosts do not seem to be important so we will ignore them from now on, although this should be understood better. In conformal gauge we can perform the following field redefinition mixing the two modes  
\bea
b\phi &=& \rho - \pi b^2 \Phi, \label{mmm}\\
b \chi &=& \rho + \pi b^2 \Phi. \label{nnn}
\eea
We rewrite the path integral now over $\rho$ and $\Phi$. We will call $\Phi$ the dilaton and define a new two-dimensional metric
\beq
g = e^{2 \rho} \hat{g}.
\eeq
Notice this metric is not the same as the worldsheet metric in the minimal string. In terms of this new metric and dilaton the minimal string can be rewritten as 
\beq
I= - \frac{1}{2} \int \sqrt{g} \left[ \Phi R + 4\mu \sinh \left( 2\pi b^2 \Phi \right) \right].
\eeq
This has the most general two-derivative form of dilaton-gravity $-\frac{1}{2} \int \Phi R + 2 U(\Phi)$, with dilaton potential given by $U(\Phi)=2\mu \sinh \left( 2 \pi b^2 \Phi \right)$ (this potential was studied for different reasons in \cite{Kyono:2017jtc, *Kyono:2017pxs, *Okumura:2018xbh}). This suggests that the minimal string is equal to a certain dilaton-gravity theory and some checks were performed in \cite{Mertens:2020hbs}. This theory simplifies in the limit $b \to 0$ where the potential is linear and the theory becomes Jackiw-Teitelboim gravity
\beq
I= - \frac{1}{2} \int \sqrt{g}~ \Phi (R +2 \Lambda),
\eeq
where in the limit we keep $\Lambda \equiv 4\pi b^2 \mu$, which becomes the absolute value of the two-dimensional cosmological constant, fixed and set to one. We will call this the pure JT gravity limit or {\it JT limit} for short. We can check that this identification is true by comparing the disk density of states \cite{Saad:2019lba}. We will extend this to the case of the minimal string deformed by tachyon operators in section \ref{sec:JTsection}. 

\subsection{String equation} \label{subsection:stringeqn}

From the perspective of the loop equations \cite{Migdal:1984gj, Eynard:2004mh}, all the information of a matrix integral is encoded in the disk density of state $\rho_0(E)$, which is also equivalent to giving the matrix potential $V(H)$. Higher genus contributions are determined from the topological recursion \cite{Eynard:2004mh,Eynard:2007kz}; for an example on how this is done see \cite{Saad:2019lba}. Therefore, if two theories that are dual to a matrix integral share the same disk density of states it means the theories are equivalent to all orders in the genus expansion. Since we will work in the double-scaling limit we will `label' the theory by $\rho_0(E)$ instead, since the precise matrix potential depends on how we regularize the theory away from the double-scaling limit. 

It will be very useful when studying deformations to look at the matrix integral from a different perspective: the string equation \cite{Brezin:1990rb,*Gross:1989vs,*Douglas:1989ve} (building on previous work \cite{Itzykson:1979fi,David:1984tx, Kazakov:1985ds,Kazakov:1985ea,Kazakov:1989bc}). To leading order in the genus expansion the string equation has the following form
\beq
\sum_{k} t_k u^k = x,
\eeq
where $x$ is a dummy variable we can use to compute certain observables and $t_k$ are the KdV couplings. We include $t_0$ on the left-hand side so at the end of the day we will always fix $x=0$. This can be derived by taking the double-scaling limit of the orthogonal polynomial method; see for example \cite{Brezin:1990rb}. Knowing $u(x)$ allows us to compute the main observable we are interested in. For example, the genus zero disk partition function is given by \cite{Banks:1989df}
\beq\label{eq:trebhse}
\left\lb {\rm Tr}\left[e^{-\beta H}\right] \right\rb_{g=0} = e^{S_0}\frac{1}{\sqrt{2\pi \beta}}\int_0^\infty dx \hspace{0.1cm} e^{-u(x)\beta}.
\eeq
The string equation provides an independent way to compute higher genus corrections. The recipe is to replace in the string equation the powers of $u$ by the Gelfand-Dickii differential operators $u^k \to R_k[u;\hbar]$ which to leading order in $\hbar\equiv e^{-S_0}$ coincide with $R_k [u] \sim u^k$ \cite{Gelfand1975,*Gelfand2}. Then one needs to solve again for $u(x;\hbar)$ and compute observables using the quantum mechanical free fermion perspective \cite{Banks:1989df} which to leading order reduces to \eqref{eq:trebhse}. This perspective is useful to derive analytical \cite{Mertens:2020hbs} and numerical results \cite{Johnson:2019eik} which would be hard to obtain from the topological recursion.

To simplify the presentation, instead of giving the KdV couplings individually we will directly define and compute the function 
\beq
\mathcal{F}(u) \equiv \sum_k t_k u^k.
\eeq
The string equation to leading order becomes $\mathcal{F}(u)=x$. The KdV coupling can be easily extracted by a Taylor expansion. We allow for the possibility of the index $k$ being unbounded, since this is needed for the JT gravity matrix integral.

The function $\mathcal{F}(u)$ fully specifies the double-scaling limit of a matrix integral. This is equivalent to giving the disk density of states which using \eqref{eq:trebhse} is related by the equation
\beq\label{eq:rhotoF}
\rho(E) =\frac{e^{S_0}}{2\pi} \int_{E_0\equiv u(0)}^E \frac{du}{\sqrt{E-u}} \frac{\partial \mathcal{F}}{\partial u},
\eeq 
where $E_0$ is the largest root of the string equation with $x=0$, given explicitly by $\mathcal{F}(E_0)=0$. This is in general a complicated equation. 

\paragraph{String equation for the $(2,p)$ minimal string:} The minimal string with $p=2m-1$ is dual to the $m$th-multicritical model of a one-matrix integral \cite{Kazakov:1989bc, Staudacher:1989fy}. In order to compare the matrix model with the results from the worldsheet CFT approach one has to turn on lower order couplings in a specific way \cite{Moore:1991ir}. We will now review the derivation of the `string equation' that matches with the CFT continuum approach \cite{Moore:1991ir, Belavin:2008kv}. This can be read off from the calculation of the disk partition function $Z(\ell)$ with fixed length boundary conditions \cite{Moore:1991ir,Fateev:2000ik}. The final answer is given by 
\beq
Z(\ell) \sim \int_0^\infty ds\hspace{0.1cm} e^{-\mu_B(s) \ell} \sinh 2\pi b s \sinh 2 \pi \frac{s}{b}.
\eeq
From this expression we can extract the disk density of states as a function of $p$ and also the cosmological constant through $\kappa$. This identifies $\mu_B(s)$ with the matrix eigenvalues $E_\mstr$ (we leave $E$ and $u$ to refer to a different convention that we specify below). The answer picking a particular normalization is given by
\beq\label{eqmsdos}
\rho(E_\mstr) = e^{S_0}\frac{p^2}{32\pi^4\kappa^2} \sinh \left(\frac{p}{2} {\rm arccosh}\left( \frac{E_\mstr}{\kappa}\right)\right).
\eeq
 Using \eqref{eq:rhotoF} we can obtain the string `string equation' associated to this theory. The result is given by
\beq\label{eqmssss}
\mathcal{F}(u_\mstr) = \frac{p^2}{32\pi^3 \sqrt{2\kappa^3}}\left[ P_{m}\left(\frac{u_\mstr}{\kappa}\right) - P_{m-2}\left(\frac{u_\mstr}{\kappa}\right)\right],
\eeq
where $\kappa \sim \sqrt{\mu}$ and $P_n(x)$ are the Legendre polynomials. We review the definition and some useful properties in Appendix \ref{appformu}. At large energies this precisely becomes the string equation of the $m$-th multicritical point of the matrix integral $u^m \sim x$, or equivalently $\rho_0(E_\mstr) \sim (\sqrt{E_\mstr})^{2m-1}$. As anticipated this answer is modified by turning on other lower couplings and near the edge this presents the universal $\rho (E_\mstr) \sim \sqrt{E_\mstr-\kappa}$ behavior.

\paragraph{JT gravity string equation:} Now we will write down the JT gravity tree-level string equation and compare it with a limit of the minimal string. It is useful to rescale and shift the matrix, and therefore its eigenvalues $E$, and the variable $u$ in the string equation correspondingly. Following \cite{Mertens:2020hbs} we define the variables $E$ and $u$ as
\beq\label{eq:MStoJTen}
E_\mstr=\kappa\Big(1+\frac{8\pi^2}{p^2}E\Big),~~~~u_\mstr=\kappa\Big(1+\frac{8\pi^2}{p^2}u \Big).
\eeq
In terms of these variables the density of states and string equation of the minimal string are independent of $\mu$ and become
\bea
\rho(E)&=&\frac{e^{S_0}}{4\pi^2} \sinh \Big(\frac{p}{2} {\rm arccosh}\Big( 1 + \frac{8 \pi^2}{p^2} E\Big)\Big) , \\
\label{eq:MSUDJT}\mathcal{F}(u)&=&  \frac{p}{16 \pi^2}\left[ P_{m}\Big(1+ \frac{8\pi^2}{p^2} u \Big) - P_{m-2}\Big(1+ \frac{8\pi^2}{p^2} u \Big)\right].
\ea 
We can now take the large $p$ limit keeping $E$ and $u$ fixed. This gives 
\beq
\rho_\jt(E)=\frac{e^{S_0}}{4\pi^2}\sinh\left(2\pi \sqrt{E}\right)~~~~\leftrightarrow~~~~\mathcal{F}_\jt(u)=\frac{\sqrt{u}}{2\pi}I_1\left(2\pi \sqrt{u}\right).
\eeq
It is now evident that this coincides with the density of states (and therefore string equation) of the matrix integral dual to pure JT gravity. 

\subsection{Deformations and Belavin-Zamolodchikov string equation} \label{sec:BZse}

We now analyze deformations of the minimal string. We will review the exact string equation proposed by Belavin and Zamolodchikov \cite{Belavin:2008kv}, building on the work of Moore, Seiberg and Staudacher \cite{Moore:1991ir}. In the next section we will make a connection between these deformations of the minimal string and the conical deformations of pure JT gravity introduced in \cite{Maxfield:2020ale} and \cite{Witten:2020wvy}.

We will focus on the following type of deformations by adding a combination of tachyon operators $\sum_n \tau_n \mathcal{T}_n$ to the minimal string action, where $\tau_n$ are the couplings of each deformation. Then the action of the deformed minimal string is, writing the deformation more explicitly,
\beq\label{eq:msactdef}
I = I_{(2,p)} - \sum_{n=2}^{m-1} \tau_n \int \sqrt{\hat{g}}\hspace{0.1cm} \mathcal{O}_{1,n}\hspace{0.1cm} e^{b(1+n)\phi},~~~~n=1,\ldots,m-1,
\eeq
where $a_n=b(n+1)/2$ is tuned such that the integrand is a marginal operator. 

The outline of the derivation is the following. When computing tachyon correlation functions on the sphere, the structure is strongly constrained by the minimal model fusion rules and conformal invariance. For example the correlator $\langle \mathcal{T}_n \mathcal{T}_{n'}\rangle_{S^2}=0$ unless $n=n'$. On the other hand, given a tree-level string equation one can derive these correlators by taking derivatives with respect to the couplings $\tau_n$'s. Strikingly, solely the conditions derived from the fusion rules completely fix the string equation. The final answer is given by
\bea
\mathcal{F}(u_\mstr)&=& \frac{p}{16 \pi^2} \left(P_{m}\left(\frac{u_\mstr}{\kappa}\right) - P_{m-2}\left(\frac{u_\mstr}{\kappa}\right)\right) \nonumber\\
&&+\sum_{L=1}^\infty \sum_{n_1,\dots,n_L=1}^{m-1} \frac{1}{L!}\Pi_{i=1}^L \lambda_{n_i}  \left(\frac{16 \pi^2}{p^2} \right)^{L-1} \, P^{(L-1)}_{m-1-\sum_{i=1}^L n_i }\left(\frac{u_\mstr}{\kappa}\right), \label{eq:MSdefgSE}
\eea
where $\lambda_n \propto \tau_n$, with the prefactor determined below. We defined the $L$th derivative of the $n$th Legendre polynomial as $P_n^{(L)}(x)\equiv \partial_x^{L} P_n (x)$ \footnote{Not to be confused with the associated Legendre polynomial which will not appear in this paper.}. The sum is over a set of $L$ integers $1\leq n_i \leq m-1$ where $i=1,\ldots,L$ and $L=1,\ldots, \infty$. As explained in Appendix \ref{appformu}, the Legendre polynomial is defined such that $P_{n}(x)=0$ when $n<0$. This implies that the sums in the second term in the right-hand-side only contribute as long as $m-1-\sum_i n_i \geq 0$. This constraint comes from analyzing resonance conditions between deformations.

The relation between the parameter $\lambda_n$ associated to the deformation and the coupling in the action $\tau_n$ depends on the precise normalization of the minimal model operators. If we pick the conventional normalization we obtain
\bea
\lambda_n = \tau_n ~{\rm Leg}(n) \frac{p^2}{16 \pi^2},
\ea
where we defined the leg-factor
\beq\label{eq:legpolestandard}
{\rm Leg}(n)\equiv \frac{i^{n-1}}{2\mu} \sqrt{ \frac{\pi \mu \gamma(n b^2)}{ (\pi \mu \gamma(b^2))^n}} ~ \frac{\Gamma(\frac{1}{b^2} - 1)}{\Gamma(\frac{1}{b^2}-n)},
\eeq
with $\gamma(x)\equiv \Gamma(x)/\Gamma(1-x)$. This relation comes from comparing the correction to the disk partition function to linear order in the deformation computed from the string equation or from the worldsheet CFT description. We do this in Appendix \ref{appendix:onept}. The leg-factor appears as well in computing sphere correlators. For example the normalized sphere two-point function is 
 \beq\label{eq:norms22ptcan}
\lb \mathcal{T}_n \mathcal{T}_{n'} \rb_{S^2} = \delta_{n'n} \frac{(2m-3)(2m-1)(2m+1)}{2m-2n-1} {\rm Leg}(n)^2.
\eeq
If we want to shift normalization for the minimal model operators by $\mathcal{O}_{1,n} \to N_n \mathcal{O}_{1,n}$ this can be achieved by shifting ${\rm Leg}(n) \to N_n {\rm Leg}(n) $. We will use this in section \ref{sec:expnorm} to study these expressions in a normalization more natural for the time-like Liouville description of the minimal model.

Now we will analyze several special cases of these formulas in order to gain some intuition. To simplify we will consider the case of a single deformation with parameter $n$ and look at various cases:

\paragraph{$\underline{n=m-1}$:} This corresponds to the case of deforming the action by the operator with the lowest dimension in the minimal model. The level of the Legendre polynomial appearing in the string equation is $m-1-\sum_i n_i$. Therefore whenever an operator with $n=m-1$ is present, it will only contribute by itself to linear order exactly. Any other term vanishes. Then its contribution to the string equation is 
\beq \label{eq:MSDefConst}
\delta \mathcal{F} \sim \lambda_{m-1} P_{0}\left(\frac{u_\mstr}{\kappa} \right) = \lambda_{m-1}.
\eeq
This is equivalent to shifting $x\to x-\lambda_{m-1}$ when computing observables. Therefore this gives an interpretation of the parameter $x$ as a specific coupling in the minimal string.

\paragraph{$\underline{\frac{m-1}{2}\leq n < m-1}$:} In this case the string equation also simplifies as long as $n>n_\star$ with $n_\star \equiv (m-1)/2$. In this range, the string equation is linear in the $\lambda_n$'s. This happens because the resonance conditions for these operators are very limited and there are no non-linear ambiguities due to contact terms \cite{Belavin:2008kv}. The solution is given by
\beq\label{MSdefshSE}
\mathcal{F}(u_\mstr)=\frac{p}{16 \pi^2}\left[ P_{m}\left(\frac{u_\mstr}{\kappa}\right) - P_{m-2}\left(\frac{u_\mstr}{\kappa}\right)\right] + \sum_{n=n_\star}^{m-1} \lambda_{n}  P_{m-n-1}\left( \frac{u_\mstr}{\kappa} \right). 
\eeq 
As $n$ decreases from $n=m-1$ to $n_\star$, the order of the polynomial goes from order zero to order $(m-1)/2$. 

\paragraph{$\underline{n < \frac{m-1}{2}}$:} This is the most complicated range. As becomes clear from the string equation, deformations with lower $n$ involve higher orders in $\lambda_n$. In this case there is no simplification and we need to consider the full expression. 

\paragraph{$\underline{n =1}$:} Finally, there is a simplification when we include operators with $n=1$. This is the identity operator and we should reproduce a shift of the cosmological constant by $-\tau_1$. In this case the bound on the terms that can appear in the string equation is $m-1-\sum_i n_i=m-1-L>0$, and therefore $L\leq m-1$. The sum can be explicitly done using results in Appendix \ref{appformu} as
\bea
\mathcal{F}(u_\mstr) &\sim& \frac{P_m(u_\mstr/\kappa)-P_{m-2}(u_\mstr/\kappa)}{2m-1}+ \sum_{n=1}^{m-1} \frac{\lambda^n}{n!} \left(\frac{16 \pi^2}{p^2} \right)^n  P^{(n-1)}_{m-n-1}(u_\mstr/\kappa) \\
&\sim& P_m\left(\frac{u_\mstr}{\kappa \sqrt{1-2\tau_1  {\rm Leg}(1)}}\right)-P_{m-2}\left(\frac{u_\mstr}{\kappa \sqrt{1-2\tau_1 {\rm Leg}(1)}}\right),
\ea
where we omit a $u$ independent prefactor that will not be important, and in the second line we have used $\lambda \hspace{.05cm} \frac{16 \pi^2}{p^2} = \tau_1 \hspace{.05cm} {\rm Leg}(n=1)$. We see that this gives back the undeformed minimal string equation with the replacement $\kappa \to \kappa \sqrt{1-2\tau_1 \hspace{.05cm} {\rm Leg}(1)}$. Using the minimal model normalization of operators we obtain ${\rm Leg}(n=1)=\frac{1}{2\mu}$. Therefore we find that the shift of $\kappa\sim \sqrt{\mu}$ is equivalent to shifting the cosmological constant $\mu \to \mu - \tau_1$, consistent with equation \eqref{eq:msactdef}.

\paragraph{} After these clarifications we see the behavior of the deformed minimal string strongly depends on whether the parameter $n$ is bigger or smaller than $n_\star$. This is reminiscent of the differences between sharp or blunt defect deformations with either $\alpha<1/2$ and $\alpha>1/2$. In the next section we will see this is not a coincidence.

\section{JT gravity with defects from the minimal string}\label{sec:JTsection}
In the previous section we studied the exact solution for deformations of the $(2,p)$ minimal string. In this section we will show that the large $p$ limit of these theories corresponds to JT gravity with a gas of defects. We will use this correspondence to find the disk density of states for JT gravity in the presence of general defects with an arbitrary defect angle.

\subsection{Sharp defects from the minimal string}
Having explained the exact solution of the deformed minimal string in section \ref{sec:BZse} we will now begin to take the JT gravity limit, namely $p\to\infty$. As a first clarification, we will take the energy and $u$ to scale as 
\beq\label{eq:MStoJTen2}
E_\mstr=\kappa\Big(1+\frac{8\pi^2}{p^2}E\Big),~~~~u_\mstr=\kappa\Big(1+\frac{8\pi^2}{p^2}u \Big),
\eeq
where $E$ and $u$ are kept fixed in the $p\to \infty$ limit.

As we take the large $p$ limit, we need to choose a scaling for both the index $n$ labeling the operator and the coupling $\tau_n$. We choose the following scaling
\beq \label{eqn:JTscaling}
n =\frac{p}{2} (1-\alpha),~~~ \tau_n = \lambda \frac{16 \pi^2}{p^2} \frac{1}{{\rm Leg}(n)},
\eeq
where we keep $\alpha$ and $\lambda$ fixed in the JT limit. The rationale for this choice will be motivated and explained below.

\begin{figure}
    \centering
    \includegraphics[scale=1]{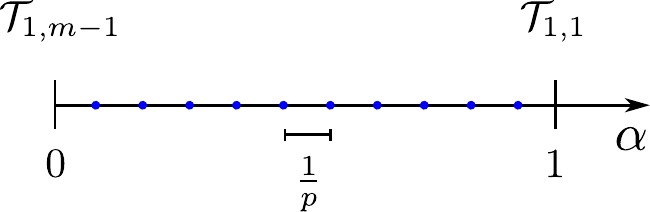}
    \caption{The relationship between the minimal string operators $\mathcal{T}_{n}$ and the JT defect parameter $\alpha$. The spacing between the operators in $\alpha$ is discrete and of order $1/p$; as we take the JT limit by sending $p$ to infinity, $\alpha$ becomes a continuous parameter.}
    \label{fig:minmodops}
\end{figure}

At finite values of $p$, $\alpha$ is a discrete parameter since $n$ is discrete and takes value in the set $\alpha \in \frac{2}{p}\cdot \mathbb{Z}$. As we take $p \to \infty$ this parameter becomes continuous. Since $n$ in the minimal string is in the range $1\leq n \leq m-1$, the parameter $\alpha$ lies between $0<\alpha<1$. The case $n=1$ corresponds to $\alpha=1-\frac{2}{p}\to_{p\to\infty}1$, while $n=m-1$ corresponds to $\alpha=\frac{1}{p}\to_{p\to\infty}0$. We give a diagram showing the relation between $n$ and $\alpha$ in figure \ref{fig:minmodops}. We saw the deformations in the minimal string changes drastically at the threshold $n_\star$. In terms of $\alpha$ this is given by $\alpha_\star=\frac{1}{2} + \frac{1}{2p} \to_{p\to\infty}\frac{1}{2}$. The string equation simplifies when $n_\star <n $ or equivalently $0<\alpha<1/2$. This is precisely the same range corresponding to sharp defects studied in \cite{Maxfield:2020ale, Witten:2020wvy}. 

We begin by considering the large $p$ limit restricted to $n_\star<n$ or $\alpha<1/2$. We write the string equation \eqref{MSdefshSE} in terms of variables that we keep fixed in the JT limit. Deformations of the string equation in this range are exactly linear so we can consider one defect without loss of generality. The finite $p$ string equation perturbed by one defect is
\beq\label{MSdefshSE2}
\mathcal{F}(u)=\frac{p}{16 \pi^2}\left[ P_{m}\left(\frac{u_\mstr}{\kappa}\right) - P_{m-2}\left(\frac{u_\mstr}{\kappa}\right)\right] + \lambda ~P_{m-\frac{p}{2}(1-\alpha)-1}\left( \frac{u_\mstr}{\kappa} \right). 
\eeq 
Using identities from Appendix \ref{appformu} the large $p$ limit with $\alpha$ and $\lambda$ fixed  gives the following answer  
\bea \label{eqn:JTstringeqnsharp}
\mathcal{F}(u) \to \frac{\sqrt{u}}{2\pi}I_1\left(2\pi \sqrt{u}\right) + \lambda\hspace{0.1cm} I_{0}\left(2\pi \alpha \sqrt{u}\right) ,
\eea
which precisely coincides with the string equation of JT gravity with a gas of sharp defects \cite{Maxfield:2020ale,Witten:2020wvy}! In that context the gas of defects are characterized by a deficit angle $2\pi(1-\alpha)$ (with $\alpha$ related to the minimal model operator) and a weighing factor $\lambda$ (related to the coupling $\tau$). It is easy to extend this to the case of multiple defect species from \eqref{MSdefshSE} 
\bea \label{eqn:JTstringeqnsharp222}
\mathcal{F}(u) \to \frac{\sqrt{u}}{2\pi}I_1\left(2\pi \sqrt{u}\right) + \sum_{i}\lambda_i\hspace{0.1cm} I_{0}\left(2\pi \alpha_i \sqrt{u}\right),
\eea
obtaining again a match with JT gravity. 

We have shown that the minimal string deformed by operators $n_\star < n$ at large $p$ becomes JT gravity with sharp defects $\alpha<1/2$. We have only compared tree-level string equations, but we would like to stress that this is enough to argue that both theories are the same to all orders in perturbation theory. This is because both theories are dual to a matrix integral, and all observables are determined by the matrix potential which can be determined from the tree-level disk density of states. Of course, this is true up to corrections not captured by the double-scaling limit.

\subsubsection*{Another check}
The upshot of the previous calculation is that the insertion of a defect from the dilaton-gravity perspective is equivalent to a tachyon vertex operator insertion in the minimal string. It is instructive to check this more explicitly in a simple example, taken from \cite{Mertens:2019tcm, Mertens:2020hbs}. We compute the expectation value of the disk partition function with a single tachyon insertion and fixed boundary length $\ell$. This is easy to do using Liouville CFT techniques and gives the answer 
\beq
\lb \mathcal{T}_{n} \rb \sim \int_0^\infty ds~e^{-\mu_B(s)\ell } \cos \left(4\pi P_n s\right),
\eeq
where $P_n=\pm i (1-nb^2)/(2b)$ is the Liouville momentum of the gravitational dressing associated to $\mathcal{O}_{1,n}$ with $n = (1-\alpha)/b^2$. Replacing this value and writing the answer in terms of $\alpha$ gives 
\beq\label{swqweqwe}
\lb \mathcal{T}_{n} \rb \sim \int_0^\infty ds e^{-\ell \kappa \cosh\left(2 \pi b s \right)} \cosh \left( \frac{2 \pi \alpha s}{b} \right).
\eeq
To take the JT limit of the result written in this form we set $s = b k$ and $\ell = \beta/(2 \pi^2 \kappa b^4)$ and take $b \rightarrow 0$ while keeping $\beta$ and $k$ fixed
\beq
\lb \mathcal{T}_{n} \rb \sim \int_0^\infty dk e^{-\beta  k^2} \cosh \left(2 \pi \alpha k \right) \sim \frac{1}{\sqrt{\beta}} e^{\frac{\pi^2 \alpha^2}{\beta}}.
\eeq
This is precisely, for a specific choice of units, the path integral of JT gravity on the disk with fixed renormalized length $\beta$. This check is valid for any $0<\alpha<1$. 

In the small $b$ limit these are heavy operators. This match can then be understood from the perspective of the semiclassical evaluation of the Liouville path integral with one heavy insertion and fixed boundary length. The classical solution is precisely a hyperbolic metric with one conical defect at the operator insertion. For a review in the present context see Appendix B.1 of \cite{Mertens:2020hbs}.

\subsubsection*{Cases with $E_0=0$}
In analogy with an analysis in \cite{Witten:2020wvy}, we can study deformations of the minimal string with $n_\star < n$ that give $E_0=0$. This requires at least two different defects and in general $\sum_n \lambda_n =0$. In this case we can analytically compute the density of states.

Using the equation for the density of states \eqref{eq:rhotoF} with the string equation \eqref{MSdefshSE2} in terms of $u$ variables, the integral identity in \eqref{appeqidintdp} immediately gives
\bea
e^{-S_0}\rho(E) &=& \frac{1}{4\pi^2} \sinh\left( \frac{p}{2}{\rm arccosh} \left(\frac{E_\mstr}{\kappa}\right)\right) \nonumber\\
&&+  \sum_{n=n_\star}^{m-1} \lambda_n  \frac{2\kappa}{p\sqrt{E_\mstr^2-\kappa^2}}\cosh\left(\frac{p-2n}{2} {\rm arccosh}\left(\frac{E_\mstr}{\kappa}\right) \right),
\ea
where we have written the final result in terms of $E_\mstr$ for simplification purposes. It is surprising from the worldsheet CFT perspective of the minimal string that all higher order corrections in $\lambda_n$ vanish! This is a nontrivial consequence of the Belavin-Zamolodchikov analysis that deserves further study. A similar observation was made in \cite{Witten:2020wvy} in the context of JT gravity. To make the connection we take the large $p$ limit of this expression giving 
\beq
e^{-S_0}\rho(E) \to \frac{1}{4\pi^2} \sinh\left( 2\pi \sqrt{E} \right) + \sum_{i} \lambda_i \frac{\cosh\left(2\pi \alpha_i \sqrt{E} \right)}{2\pi \sqrt{E}}.
\eeq
This is precisely the density of states of JT gravity with sharp defects for $E_0=0$.

\subsection{Solution with general defects}

\subsubsection*{One defect species}
The most general string equation for an arbitrary deformation is complicated, so we will begin by analyzing a simpler case where only one deformation $\tau_n$ is turned on with $n<n_\star$. Then \eqref{eq:MSdefgSE} simplifies into the following equation 
\beq\label{eq:MSdefgSEone}
\mathcal{F}(u)=\frac{p}{16 \pi^2}\left[ P_{m}\left(\frac{u_\mstr}{\kappa}\right) - P_{m-2}\left(\frac{u_\mstr}{\kappa}\right)\right] + \sum_{L=1}^{\lfloor \frac{m-1}{n}\rfloor} \frac{{\lambda_{n}^L}}{L!} \left(\frac{16 \pi^2}{p^2} \right)^{L-1}  P^{(L-1)}_{m-Ln-1}\left(\frac{u_\mstr}{\kappa}\right),
\eeq
where we remind the reader that $P^{(L)}_n(x)\equiv \partial_x^L P_n(x)$ is the derivative of the Legendre polynomial. The new feature compared to \eqref{MSdefshSE} is the fact that now the string equation is non-linear as a function of the deformation. This modification will survive the large $p$ limit. 

Now we take the JT limit. In order to do this we take $p\to\infty$, or equivalently $m\to\infty$, while scaling the index of the minimal model operator as $n=p(1-\alpha)/2$ and the coupling as in \eqref{eqn:JTscaling}. This scaling gives JT gravity with a gas of defects with weight $\lambda$ and defect angle $\alpha$. The final answer for the string equation obtained from \eqref{eq:MSdefgSEone}, using identities in Appendix \ref{appformu}, is given by 
\bea
\mathcal{F}(u)&=&\frac{\sqrt{u}}{2\pi} I_{1}\left( 2 \pi \sqrt{u} \right) + \lambda I_{0}\left(2\pi \alpha \sqrt{u} \right)\nonumber\\
&&+\sum_{L=2}^{\lfloor \frac{1}{1-\alpha} \rfloor} \frac{\lambda^L}{L!} \left( \frac{2\pi(1-L(1-\alpha))}{\sqrt{u}} \right)^{L-1} I_{L-1}\left( 2\pi (1-L(1-\alpha))\sqrt{u} \right). \label{eq:onegdef} 
\eea
The first line contains the two terms that already appear in the string equation for sharp defects. In the second line we show the new terms that appear when $\alpha>1/2$. The nonlinear terms of order $\lambda^L$ appear only if $L \leq \lfloor \frac{1}{1-\alpha} \rfloor$. This has a geometric interpretation since there is a bound on the number of individual defects that can be merged into a single defect. Moreover, the combination appearing in each term, $2\pi(1-L(1-\alpha))$ is precisely the deficit angle after merging $L$ defects of angle $2\pi(1-\alpha)$. It would be nice to understand how to derive this formula directly from the perspective of JT gravity but we leave this for future work.

Finally, using this tree-level string equation we can find the density of states to leading order in the genus expansion. Using \eqref{eq:rhotoF} this is given by 
\beq \label{JTDefDiskDOS}
\rho(E) =\frac{e^{S_0}}{2\pi}\sum_{L=0}^{\lfloor \frac{1}{1-\alpha} \rfloor} \frac{\lambda^L}{2L!} \int_{E_0}^{E} \frac{du}{\sqrt{E-u}} \left( \frac{2\pi(1-L(1-\alpha))}{\sqrt{u}} \right)^{L} I_{L}\left( 2\pi (1-L(1-\alpha))\sqrt{u} \right).
\eeq
The edge of the spectrum $E_0$ appearing in this expression is defined as the largest root of \eqref{eq:onegdef} solving $\mathcal{F}(u_0=1+\frac{8\pi^2}{p^2}E_0)=0$. For $\alpha \leq 1/2$ this coincides with the answer found in \cite{Maxfield:2020ale, Witten:2020wvy} and generalizes it to $\alpha>1/2$. Contrary to previous expectations this is a non-analytic function of $\alpha$ and the result changes drastically as $\alpha$ approaches one.

\subsubsection*{General case}
We now generalize the previous discussion to an arbitrary number of defect species put together. In order to do this we take the JT gravity limit of the most general deformed minimal string equation \eqref{eq:MSdefgSE}. The calculation is very similar to what we have already done and therefore we give the final answer
\bea
\mathcal{F}(u)=  \sum'_{\{ \ell_i\}} \frac{\prod_i \lambda_i^{\ell_i}}{L!} \left( \frac{2\pi(1-\sum_i \ell_i(1-\alpha_i))}{\sqrt{u}} \right)^{L-1} I_{L-1}\Big( 2\pi \Big( 1-\sum_i \ell_i (1-\alpha_i)\Big)\sqrt{u} \Big),\label{JTdefgeneralsemul}
\eea
where $L=\sum_i\ell_i$ and the sum includes all permutations of $\lambda_i$ for a particular configuration $\{ \ell_i\}$. The case with all $\ell_i=0$ and therefore $L=0$ gives back the pure JT gravity contribution. We see this is a simple generalization of the one species case \eqref{eq:onegdef}. The prime in the summation means that we only include configurations $\{\ell_i\}$ such that the following identity is satisfied
\beq
1-\sum_i \ell_i (1-\alpha_i)>0.
\eeq
This bound has a simple geometric interpretation. It specifies the maximum number of defects $\alpha_i$ that can merge together, and it is a straightforward generalization of the bound on $L$ appearing in \eqref{eq:onegdef}.

\subsubsection*{Structure of the perturbative expansion} 
We now explain general features of the perturbative expansion in $\lambda$ of the partition function, and we relate it to some geometrical intuition developed in section \ref{sec:generaldefects}. For simplicity we focus on the case of one defect species. 

To begin consider the case of sharp defects. For $0<\alpha \leq 1/2$ the first orders in the expansion are 
\beq
\alpha\leq\frac{1}{2}: ~~~~Z_{\jt} = e^{S_0}\frac{e^{\frac{\pi^2}{\beta}}}{4\sqrt{\pi} \beta^{3/2}},~~~Z_{1-\df} =e^{S_0}\frac{e^{\frac{\pi^2 \alpha^2}{\beta}}}{2\sqrt{\pi \beta}},~~~Z_{2-\df}=e^{S_0} \frac{\sqrt{\beta}}{2\sqrt{\pi}},~~~\ldots,
\eeq
where the dots indicate contribution with larger number of defects. The exponential appearing in $Z_{\jt}$ and $Z_{1-\df}$ come from classical solutions, while the prefactor comes from perturbative quantum corrections. On the other hand for more than one defect there is no classical solution. This manifests itself in the fact that for $Z_{k-\df}$ with $k>1$ the answer is always given by a finite polynomial in $\sqrt{\beta}$ (it is easy to show this property using the string equation).

Consider now the case $1/2<\alpha<2/3$ for which the maximum order in $\lambda$ in the string equation is $\lfloor \frac{1}{1-\alpha} \rfloor=2$. The perturbative expansion now has new terms 
\bea
\frac{1}{2}<\alpha<\frac{2}{3}: ~~&&Z_{\jt} =e^{S_0} \frac{e^{\frac{\pi^2}{\beta}}}{4\sqrt{\pi} \beta^{3/2}},~~~Z_{1-\df} =e^{S_0}\frac{e^{\frac{\pi^2\alpha^2}{\beta}}}{2\sqrt{\pi \beta}},~~~Z_{2-\df} =e^{S_0}\frac{\sqrt{\beta}}{2\sqrt{\pi}} e^{\frac{\pi^2(1-2\alpha)^2}{\beta}} ,\nonumber\\
&&Z_{3-\df} =e^{S_0}\frac{\pi^{3/2}}{6 \sqrt{\beta}}(\beta (12-6\alpha(4-3\alpha))-\pi^2),~~~\ldots .
\ea
The first two terms corresponding to zero or one defect are unchanged. This is true for defects in the whole range $0<\alpha<1$. But now for $1/2<\alpha<2/3$ we see the two-defect term gets modified and a new exponential term appears. This is consistent with the analysis in section \ref{sec:generaldefects}, in particular the exponent of $Z_{2-\df}$ is precisely the same as the classical contribution of two defects of angle $\alpha$ merging into a single one of angle $2\pi(2\alpha-1) > 0$. The contribution of $Z_{k-\df}$ for all $k>2$ can be shown to be given by a finite sum of powers, consistent with the fact that for $1/2<\alpha<2/3$ there is no classical solution with three or more merged defects\footnote{For $1/2<\alpha<2/3$ and $k>2$ there is a geodesic in the geometry we can use to glue with trumpets. Nevertheless in this case a different calculation of the WP volume is required \cite{tan2004generalizations}. Therefore, more generally contributions with $Z_{k-\text{def}}$ and $k>\lfloor \frac{1}{1-\alpha} \rfloor$ do not need to match with the ones from sharp defects continued to values of $\alpha>1/2$, and we find they are different.}. 

A similar structure is valid in the whole range $0<\alpha<1$ consistent with section \ref{sec:generaldefects}. If we look at the high temperature limit of small $\beta$, then the first terms have the behavior \beq
Z_{L-\df} \sim e^{S_0}\beta^{L-\frac{3}{2}} e^{\frac{\pi^2(1-L(1-\alpha))^2}{\beta}},~~~\text{for }~~L \leq \Big\lfloor \frac{1}{1-\alpha} \Big\rfloor,
\eeq
while for $Z_{k-\df}$ with $k>L$ the answer is always a finite sum of powers. It would be interesting to understand these results directly from a JT gravity with defects path integral calculation, but we leave this for future work. Some progress in this direction was done in \cite{Budd}. We collect more results regarding the perturbative expansion in Appendix \ref{app:pertexpexamples}.

\subsection{Summary: Types of defects}
In this section we summarize the different types of defects and their correspondence to the minimal string deformations outlined at the end of section \ref{sec:BZse}.

\paragraph{\underline{$\alpha=0$:}} This gives a gas of cusps, and is obtained in the large $p$ limit of the $\mathcal{T}_{m-1}$ deformation. The string equation in this case is simply 
\bea
\mathcal{F}(u) =\frac{\sqrt{u}}{2\pi}I_1\left(2\pi \sqrt{u}\right) + \lambda.
\eea
Just like the $\mathcal{T}_{m-1}$ deformation, this is equivalent to shifting the variable $x\to x-\lambda$ in the string equation and gives a geometric interpretation for this dummy variable used to compute expectation values. From the dilaton-gravity perspective this amounts to adding a term to the dilaton potential proportional to $e^{-2\pi \Phi}$.

\paragraph{\underline{$0<\alpha\leq1/2$:}} These are the sharp defects studied in \cite{Maxfield:2020ale,Witten:2020wvy} and they are obtained from deformations between $\mathcal{T}_{n_\star},\ldots, \mathcal{T}_{m-2}$. The string equation is exactly linear in the deformation parameter $\lambda$
\bea \label{eqn:JTstringeqnsharp222}
\mathcal{F}(u) \to \frac{\sqrt{u}}{2\pi}I_1\left(2\pi \sqrt{u}\right) + \lambda\hspace{0.1cm} I_{0}\left(2\pi \alpha \sqrt{u}\right) .
\eea
For $\alpha=1/2$, the contribution with $L=2$ is exactly zero\footnote{This happens whenever $(1-\alpha)^{-1}$ is an integer since then the contribution from $L_{\rm max}=\big\lfloor \frac{1}{1-\alpha} \big\rfloor$ is proportional to $(1-L_{\rm max}(1-\alpha))^{L_{\rm max}-1}=0$.}. This was implicitly assumed in \cite{Maxfield:2020ale} since this case is relevant to the application to 3D, and here we can verify this.

\paragraph{\underline{$1/2<\alpha<1$:}} They correspond to blunt defects, for which the SSS recipe cannot be applied. The string equation is non-linear in the deformation
\bea
\hspace{-0.5cm}\mathcal{F}(u)\to \sum_{L=0}^{\lfloor \frac{1}{1-\alpha} \rfloor} \frac{\lambda^L}{L!} \left( \frac{2\pi(1-L(1-\alpha))}{\sqrt{u}} \right)^{L-1} I_{L-1}\left( 2\pi (1-L(1-\alpha))\sqrt{u} \right),
\eea
and we propose the new terms are related to the possibility of defects merging and new classical solutions compared with sharp defects.

\paragraph{\underline{$\alpha=1$:}} The deformation with $\alpha=1$ corresponds to the large $p$ limit of the deformation $\mathcal{T}_1$, which in the minimal string simply shifts the cosmological constant. From the JT perspective, when $\alpha=1$ we have a gas of insertions with vanishing deficit angle and therefore we expect to recover pure JT gravity. Indeed this is the case:
\bea
\mathcal{F}(u) &\to&\frac{\sqrt{u}}{2\pi} I_{1}\left( 2 \pi \sqrt{u} \right)+ \sum_{L=1}^{\infty} \frac{\lambda^L}{L!} \left( \frac{2\pi}{\sqrt{u}} \right)^{L-1} I_{L-1}\left( 2\pi \sqrt{u} \right)\\
&=&\frac{\sqrt{u+2\lambda}}{2\pi} I_{1}\left( 2 \pi \sqrt{u+2\lambda} \right).
\eea
This is precisely the JT gravity string equation up to a simple shift of $u\to u+2\lambda$. The density of states associated to this is simply
\beq\label{eq:JTdefAlphaEqOne}
\rho(E) = \frac{e^{S_0}}{4\pi^2}\sinh\left(2\pi\sqrt{E-E_0}\right),~~~E_0=-2\lambda.
\eeq
In section \ref{sec:dilgrav}, we give a possible interpretation of this shift from the dilaton-gravity perspective. It is interesting that the $\alpha\to1$ limit of the partition function in the disk with a single defect does not give back pure JT gravity. Instead we need to sum over a gas of points with zero deficit angle in order to recover the undeformed theory.

\subsection{Defect generating function}

We have presented the solution to JT gravity with a gas of generic defect species parameterized by their weight $\lambda_i$ and angle $\alpha_i$. It will be convenient to recast this information about the theory in the following defect generating function $W(y)$. This is defined as
\beq\label{eq:defgenfunc}
W(y) \equiv \sum_i \lambda_i e^{-2\pi (1-\alpha_i)y}.
\eeq
Characterizing the species present is equivalent to giving the function $W(y)$, with some restriction on $W$ coming from $0<\alpha_i<1$. 

The solution found through the string equation \eqref{JTdefgeneralsemul} is not very transparent when describing the spectrum of defects in terms of $W(y)$. As we show in Appendix \ref{app:eqLTSE}, using some integral identities for Bessel functions, the string equation \eqref{JTdefgeneralsemul} can be exactly rewritten as an inverse Laplace transform\footnote{We thank T. Budd for pointing this out \cite{Budd}.} 
\beq \label{eq:invlapse}
\mathcal{F}(u)=\int_{\mathcal{C}} \frac{dy}{2\pi i}e^{2\pi y} \left(y-\sqrt{y^2 -u - 2 W(y)} \right),
\eeq
which now depends on the function $W(y)$ in a very simple way. The contour is along the imaginary axis with all singularities to the left. 

Moreover, this expression can be inserted in \eqref{eq:rhotoF} to obtain a general formula for the disk density of states as a function of the defect generating function $W(y)$ 
\beq \label{eq:invlapsedos}
\rho(E) = \frac{e^{S_0}}{2\pi}\int_{\mathcal{C}} \frac{dy}{2\pi i} \,e^{2\pi y}\, \tanh^{-1} \left(\sqrt{\frac{E-E_0}{y^2-2W(y)-E_0}} \right),
\eeq
where the edge of the spectrum $E_0$ can be found by solving $\mathcal{F}(E_0)=0$. From these expressions it becomes evident that adding a gas of defects with angle $\alpha \to 1$ has the effect of shifting the generating function by a $y$-independent constant. From \eqref{eq:invlapse} we see such a shift $W(y)\to W(y) + c$, for some constant $c$, can be absorbed by a shift $u\to u-2c$, or equivalently a shift in the energy $E\to E-2c$. We have seen this explicitly in a simpler case at the end of the previous section.

These expressions will be extremely useful in the next section when we reinterpret this theory as a solution of 2D dilaton-gravity.

\section{Dilaton-gravity}\label{sec:dilgrav}

We will argue that there is a connection between deformations of the minimal string and dilaton-gravity theories, and we study the large $p$ limit of these theories. The minimal string formulation implies a precise relation between the defect parameters and the dilaton potential which differs from the one proposed in \cite{Witten:2020wvy}. We then use the Belavin-Zamolodchikov string equation to propose an exact solution of these dilaton-gravity theories.

\subsection{The minimal string as 2D dilaton-gravity}\label{sec:expnorm}

To explain the first point we again use the argument of \cite{StanfordSeiberg} (see also \cite{Mertens:2020hbs}) to rewrite the minimal string action in terms of a time-like Liouville field. We apply the same field redefinition \eqref{mmm}-\eqref{nnn} to the tachyon insertions present in the deformation of the minimal string action. Ignoring changes in normalization, this gives
\beq
\int \sqrt{\hat{g}}~\mathcal{O}_{1,n}~ e^{2\alpha_L \phi}\to \int \sqrt{\hat{g}}~e^{2\alpha_M \chi}~ e^{2\alpha_L \phi} \to \int \sqrt{g} ~e^{-2\pi b^2 n \Phi},
\eeq
where $g$ is the JT gravity metric and $\Phi$ the JT dilaton. The final proposal is that the deformed minimal string is equivalent to a two-dimensional dilaton-gravity theory
\beq
I= - \frac{1}{2} \int \sqrt{g} \left[ \Phi R + 2 U(\Phi) \right],
\eeq
with the following dilaton potential 
\beq\label{eq:dilpotdefms}
U(\Phi) = 2\mu \sinh \left( 2\pi b^2 \Phi \right) +  \sum_{n=1}^{m-1}  \tau_n \hspace{0.1cm} e^{-2\pi b^2 n \Phi},
\eeq
where $n=1,\ldots, m-1$ \footnote{It would be interesting to understand from the dilaton-gravity perspective whether the exponent in the dilaton potential has to be quantized at finite $p$. A more complete understanding of the minimal model as time-like Liouville would very likely answer this question (see \cite{KapecMahajan} for some progress in this direction).}. The first term is the undeformed minimal string potential derived in section \ref{sec:minstringasdilatongrav}. We can take the JT limit of this dilaton-gravity action using the scaling introduced in \eqref{eqn:JTscaling} $n=(1-\alpha)/b^2$. Each deformation term becomes 
\beq
\tau_n\int \sqrt{g}~e^{-2\pi b^2 n \Phi} \to \tau_n \int \sqrt{g}~e^{-2\pi (1-\alpha) \Phi},
\eeq
which is the same dilaton potential associated to one defect species. This gives yet another perspective on why deformations of the minimal string matched with JT gravity with defects in the previous section. 

The conventional normalization of the minimal model operator does not match with the time-like Liouville exponential required in this derivation. Therefore the parameter $\tau_n$ here is rescaled with respect to the one used in the previous section. We analyze this in detail in the next section. 

\subsubsection*{Minimal string normalization}

The Belavin-Zamolodchikov string equation, which gives the exact solution of the theory, is written in \eqref{eq:MSdefgSE} in terms of $\lambda_n$, related to the coupling in the dilaton potential $\tau_n$ by
 \be
\lambda_n = \tau_n \hspace{0.1cm}\frac{p^2}{16\pi^2} \hspace{0.1cm}{\rm Leg}(n).
\ee
 The leg-factor is defined in \eqref{eq:norms22ptcan} and depends on the precise normalization of the minimal model operator. In order to compute the dilaton potential in \eqref{eq:dilpotdefms} as a function of the $\lambda_n$'s we need to compute the leg-factor corresponding to the exponential time-like Liouville normalization, 
 \be\label{eq:eqnormexp1}
e^{2 \hat{a}_n \chi}= N^{(E)}_n \mathcal{O}_{1,n},
\ee
where $\mathcal{O}_{1,n}$ denotes the minimal model with the standard normalization used in equation \eqref{eq:legpolestandard}. We exclude the definition of the prefactor $N^{(E)}_n$ for convenience; its precise expression can be found in Appendix C of \cite{Zamolodchikov:2005fy}. Instead we quote directly the result for the leg-factor corresponding to the exponential normalization\footnote{Our definition has an extra factor of $-1/2$ relative to C.18 in \cite{Zamolodchikov:2005fy}; this is to normalize the correlation functions to be consistent with the convention of \cite{Belavin:2008kv}.}
\be
{\rm Leg}(n) = \frac{\gamma(n b^2)}{2\mu_M \gamma(-b^2)}\left(\frac{-\gamma(-b^2)}{\gamma(b^2)}\right)^{\frac{1+n}{2}}, 
\ee
where $\mu_M$ is the cosmological constant of the time-like Liouville field which we take to be $-\mu$, consistent with \eqref{eqn:timelikeaction}, and $\gamma(x) = \Gamma(x)/\Gamma(1-x)$. If we take the JT limit by sending $b\rightarrow0$ for a deformation with $n=(1-\alpha)/b^2$  we find that 
\be \label{eq:exp_coupling_JT}
\tau_n = \frac{2\pi \lambda_n}{\gamma(1-\alpha)}e^{-2(1-\alpha)c},
\ee
where $c$ is the Euler–Mascheroni constant and we have again set $4 \pi b^2 \mu = 1$ to recast the JT action in the standard form. The exponential term can be removed by a simple shift of the dilaton. 

The final result of this section is that, following the minimal string quantization, the string equation \eqref{eq:invlapse} provides an exact solution for the following 2D dilaton-gravity 
\beq\label{eq:minstrindilpot}
U(\Phi) = \Phi + \sum_{i} \frac{2\pi \lambda_i }{\gamma(1-\alpha_i)} e^{-2\pi(1-\alpha_i)\Phi},
\eeq
where the sum is over species of defects with parameters $\alpha_i$ and $\lambda_i$. The advantage of this choice is the existence of a good semiclassical limit, which we can probe in the limit $\alpha \to 1$ where the backreaction from the conical defect is small. Using that $\gamma(1-\alpha)\approx(1-\alpha)^{-1}$, the dilaton potential becomes approximately $U(\Phi) = \Phi+ \sum_{i} 2\pi(1-\alpha_i) \lambda_i e^{-2\pi(1-\alpha_i)\Phi}$. It was checked in Appendix D of \cite{Maxfield:2020ale} that this extra factor of $2\pi(1-\alpha)$ guarantees a match with the semiclassical dilaton-gravity calculation. We will come back to this in the next section where we analyze polynomial dilaton potentials.

\subsection{Polynomial potentials} \label{sec:PolyPotentials}
In this section we explain how to generate nearly polynomial dilaton potentials
\be
U(\Phi) = \Phi + \sum_{n=2}^{N} \eta_n \Phi^n + \cdots
\ee
by choosing certain combinations of defect parameters in (\ref{eq:minstrindilpot}). By nearly polynomial we mean that for some large interior region in the geometry the potential has the desired form, but near the asymptotic boundary where the dilaton is very large it reduces to JT. 

One way to construct a potential of lowest degree $m$ is to introduce $m+1$ defects with $\theta_i = 2\pi(1-\alpha_i) \approx 0$. After expanding the exponentials in $\theta_i \Phi$ the terms up to $\Phi^m$ can be cancelled by tuning the couplings $\tau_i$, with one coupling leftover to specify the free parameter $\eta_m$. The behavior of the potential is such that there is an arbitrarily large polynomial region defined by $\theta_i \Phi \ll 1$ which smoothly connects to a JT region near the asymptotic boundary where the dilaton becomes large. This ensures that the boundary is asymptotically AdS$_2$ and these theories are unambiguously defined through the string equation \eqref{JTdefgeneralsemul}.

\begin{figure}
    \centering
    \includegraphics[scale=0.5]{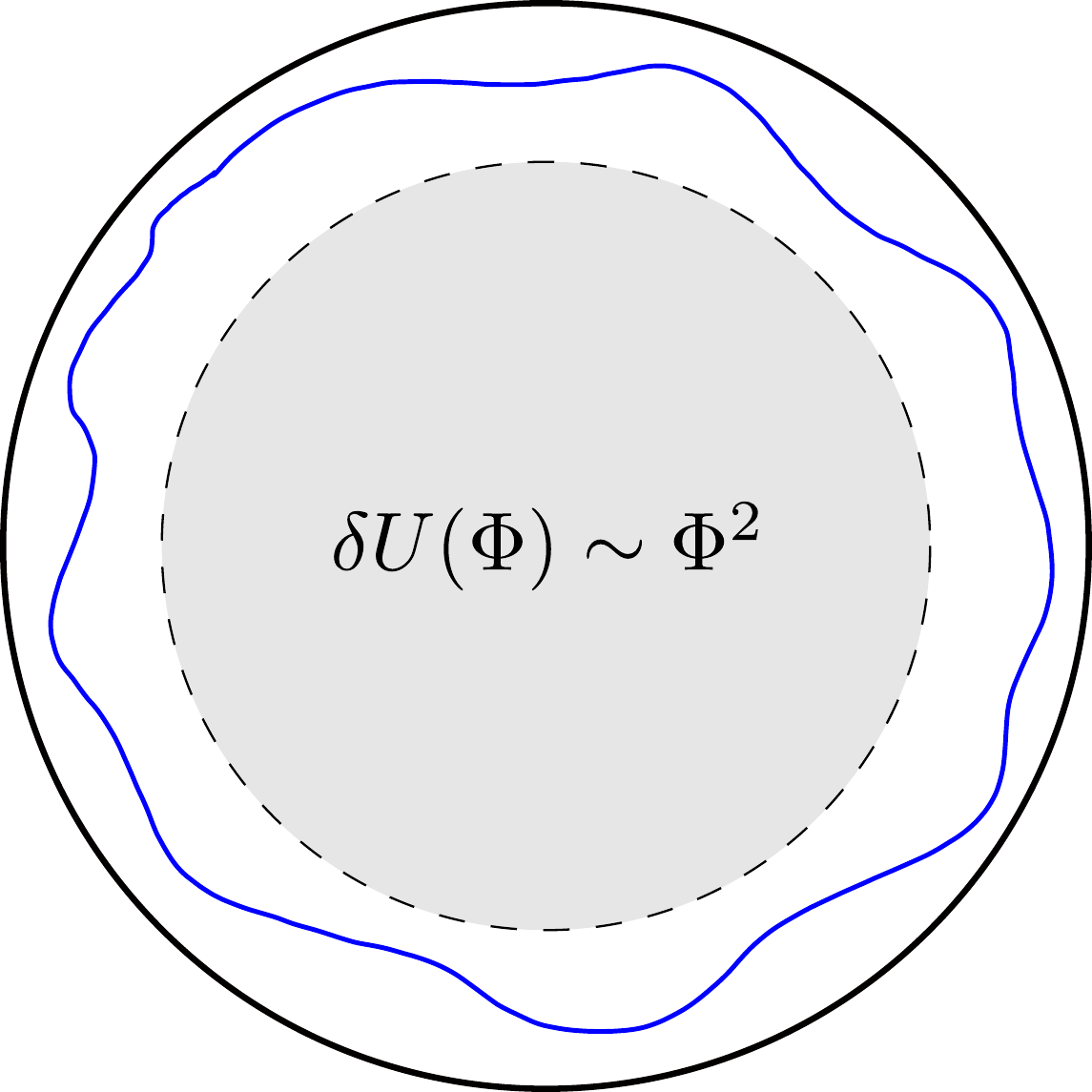}
    \caption{We illustrate the geometry for the quadratic deformation constructed in section \ref{sec:PolyPotentials}. The blue curve corresponds to the Schwarzian boundary, while the shaded region corresponds to the portion of the disk where the deformation to the dilaton potential is approximately quadratic. The region in between corresponds to the transition from the nearly polynomial behavior to AdS$_2$ asymptotics.}
    \label{fig:phisquaredDisk}
\end{figure}

We demonstrate the above procedure with a simple example of the quadratic potential $U(\Phi) = \Phi + \frac{\eta}{2} \Phi^2 + \cdots$, for which the semiclassical limit was studied in \cite{Kitaev:2017awl}.  We begin by turning on three general defects with corresponding potential
\be \label{gen3defpot}
U(\Phi) = \Phi + \tau_1 e^{-\theta_1 \Phi} + \tau_2 e^{-\theta_2 \Phi} + \tau_3 e^{-\theta_3 \Phi},
\ee
where the linear term is the undeformed JT potential. For $\theta_i \Phi \ll 1$ we can expand the exponentials to find
\be
U(\Phi) = \Phi + \tau_1 + \tau_2 + \tau_3 - (\tau_1 \theta_1 + \tau_2 \theta_2 + \tau_3 \theta_3)\Phi + \frac{1}{2}(\tau_1 \theta_1^2 + \tau_2 \theta_2^2 + \tau_3 \theta_3^2)\Phi^2 + \cdots
\ee
To get the leading order behavior of $\Phi^2$ we can choose the following parameters $\theta_1 = \theta_2/2 = \theta_3/3 \equiv \theta$ and $\tau_1 = -\tau_2/2 = \tau_3 \equiv \tau$, with $\theta$ close to zero. If we now formally take $\theta \to 0$ while keeping the combination $\eta \equiv 2 \theta^2 \tau$ fixed at a small but finite value, we arrive at the following potential
\be
U(\Phi) = \Phi + \frac{\eta}{2} \Phi^2.
\ee
The quantity that will be relevant to us will be the prepotential, which is defined by
\be \label{phi2gravity}
\widehat{U}(\Phi) = 2 \int_0^\Phi U(\Phi') d\Phi' = \Phi^2 +\frac{\eta}{3} \Phi^3.
\ee
Higher order polynomial potentials can be constructed by tuning defect parameters in a similar way.

Note that since we have constructed this potential by turning on defects this approximation is only valid when $\theta \Phi \ll 1$. At any finite value of $\theta$, this condition only holds for a region near the center of the disk. The deformation can be turned on in a large portion of the bulk by taking $\theta$ small, but there is a transition to JT asymptotics near the boundary. See figure \ref{fig:phisquaredDisk} for an illustration of this. As we will see in the next section, as long as we assume $\eta$ is fixed to be small but finite and only work perturbatively in $\eta$, we can consistently take the $\theta \to 0$ limit.

\subsection{Solution from string equation}
We now solve for the density of states from the string equation for the quadratic potential and compare to the answer from the semiclassical gravity calculation. The string equation for this theory is given by choosing the same combination of defect parameters as above in the defect generating function \eqref{eq:defgenfunc} which gives
\be \label{eq:defgenfuncquadratic}
W(y)=-\frac{\eta}{6}y^3
\ee
where we have taken the $\theta \to 0$ limit and added a defect with zero deficit angle to cancel the constant term. Examining \eqref{eq:invlapse} and \eqref{eq:invlapsedos}, we see that the combination $y^2 -2 W(y)$ appearing in the string equation can be identified with the prepotential $\widehat{U}(\Phi \to y)$, with the $y^2$ term corresponding to the JT term. To make this identification, it is important to use the quantization we obtained from the minimal string in equation \eqref{eq:minstrindilpot}. Now using either \eqref{eq:rhotoF} or \eqref{eq:invlapsedos} we can obtain the exact density of states
\be
\rho(E) = \frac{e^{S_0}}{2\pi}\int \frac{dy}{2\pi i} e^{2\pi y}~ \tanh^{-1} \left(\sqrt{\frac{E}{\widehat{U}(y)}} \right),~~~ \widehat{U}(y) = y^2 + \frac{\eta}{3} y^3
\ee
where we have used $E_0 = 0$ to all orders in $\eta$. This can now be evaluated perturbatively in $\eta$ and we find
\be \label{eq:rhoPhiSquared}
\rho(E) = \frac{e^{S_0}}{2\pi} \left( \frac{\sinh\left(2\pi \sqrt{E}\right)}{2\pi} -\frac{\eta}{6} E \sinh\left(2\pi \sqrt{E}\right) + \cdots \right).
\ee
For each term in perturbation theory in $\eta$ we do the $y$ integral along a contour in the imaginary direction with all singularities to the left, appropriate to an inverse Laplace transform. This is the prediction for the density of states from the string equation for the quadratic potential. It would be interesting to connect this solution to the quantization of the non-local boundary action derived by Kitaev and Suh \cite{Kitaev:2017awl}.

We now comment on the validity of the string equation \eqref{eq:invlapse} and density of states \eqref{eq:invlapsedos} for more general choices of $W(y)$ beyond sums of exponentials. The expression for the string equation is poorly behaved when we formally send $\theta \to 0$ to achieve \eqref{eq:defgenfuncquadratic}. The integrand is unbounded as $y$ becomes large and the branch cut structure becomes complicated. These issues are ameliorated if one works directly with the density of states \eqref{eq:invlapsedos} where the integrand is better behaved. Still working non-perturbatively in $\eta$, an issue remains regarding the choice of contour of the $y$ integral. This is not a problem in the case of defects producing exponential terms so to solve this we can go back to consider finite $\theta$. In this case the contour is uniquely defined and we expect this to indicate unambiguously which contour to use in \eqref{eq:invlapsedos}. We leave a more thorough investigation of this issue for future work. 

\subsubsection*{Check: Semiclassical limit}
We will see now that the large $E$ limit of the density of states we propose matches the semiclassical gravity calculations obtained in \cite{Kitaev:2017awl}; see also \cite{Grumiller:2007ju} and more recently \cite{Witten:2020ert}. In the semiclassical limit, the energy in terms of the prepotential is $E = \widehat{U}(\Phi_0)$ where $\Phi_0$ is the value of the dilaton at the horizon. Since the entropy is proportional to the dilaton at the horizon, the density of states is immediately given by
\be\label{eq:semicl}
\rho_{\rm classical}(E) \approx \frac{e^{S_0}}{8\pi^2} \exp\left(2\pi \hspace{0.1cm}\widehat{U}^{-1}(E)\right).
\ee
We now show that this formula matches (\ref{eq:rhoPhiSquared}) for the case of quadratic potential. Taking the large energy limit of \eqref{eq:rhoPhiSquared} and expanding for small $\eta$ we find
\beq
\log \rho(E) = S_0 + 2\pi \sqrt{E} - \frac{\eta}{3} \pi E + \mathcal{O}(\eta^2, 1/\sqrt{E})
\eeq
This expression matches with the semiclassical answer above since expanding \eqref{eq:semicl} for small $\eta$ gives $\log \rho_{\rm classical} = S_0 +2\pi \widehat{U}^{-1}(E) \approx S_0 + 2 \pi \sqrt{E} - \frac{\eta}{3} \pi E + \mathcal{O}(\eta^2)$.

Moreover, at each order in $\eta$ we can keep the terms in the exact density of states that grow fastest at large energies. We have checked these terms match with \eqref{eq:semicl} to leading order in large $E$ up to $\mathcal{O}(\eta^5)$. This suggests that our string equation correctly captures the semiclassical behavior of general dilaton potentials that can be built by turning on defect deformations. We note that this result was obtained directly from a polynomial $W(y)$ and is therefore independent of how it is regulated by a sum of exponentials in the asymptotically AdS$_2$ region.

\section{Discussion}\label{sec:discussions}

In this paper we have pointed out a connection between deformations of the $(2,p)$ minimal string by tachyon-like operators in the large $p$ limit and 2D JT gravity with a gas of defects. We also studied the connection between these theories formulated as 2D dilaton-gravities with general potentials. Using the minimal string we found an exact solution for JT gravity with defects when the defect angle is blunt, or in our notation $\alpha>1/2$. The structure of the solution is more complicated than the case of sharp defects studied in \cite{Maxfield:2020ale,Witten:2020wvy}, and we gave a geometrical interpretation of these new features. Finally we have used this solution to propose an exact disk density of states for a general class of dilaton potentials.

We finish with some open questions.

\subsection*{Path integral with general defects}

In the presence of general defects we can still integrate out the dilaton first, producing an integral over the moduli space of hyperbolic surfaces with cone points. When $\alpha<1/2$, the geometry has geodesics that can be used to cut and glue by decomposing the surface into trumpets and pairs of pants, following the recipe of \cite{Saad:2019lba}. For the case $\alpha\geq1/2$ we have used the minimal string to solve the theory. In this case there might not exist geodesics we can use to cut and glue and it would be very interesting to understand how to directly evaluate the JT path integral. Moreover, after integrating out the dilaton, we are left with an integral over the moduli space of hyperbolic surfaces with cone points, for which the corresponding Weil-Petersson volumes have only been calculated for $\alpha<1/2$ \cite{tan2004generalizations,do2006weilpetersson}. Therefore, our results can be seen as a physicist derivation of what these volumes are for more general defects. It would be interesting to understand this from first principles \cite{Budd}. 

A possible approach to solve this problem is to rewrite the JT path integral with defects as a sum over discretized surfaces with constant negative curvature. This sum over discrete surfaces cannot be done with a matrix integral, which does not fix the curvature locally. Instead, a related problem was actually studied some time ago using the model of dually weighted graphs \cite{Kazakov:1996zm} \footnote{We thank V. Kazakov for pointing this out and for several discussions.}. Matching these discrete results to the continuum approach is an open problem being pursued in \cite{FMK}.

\subsection*{Leg-factor from JT gravity}

When relating the exact solution of JT gravity with defects to the 2D dilaton-gravity formulation through the dilaton potential, some ambiguities appear. This was explained in \cite{Witten:2020wvy}. In general different renormalization procedures can lead to different quantizations of the same classical theory and 2D dilaton-gravity is no exception. 

The natural choice for JT gravity made in \cite{Witten:2020wvy} associates the classical dilaton potential with the defect generating function \eqref{eq:defgenfunc}. In this paper we have studied a different quantization provided by the minimal string, recast as a time-like Liouville coupled to a standard gravitational Liouville theory. This gives a different identification between the defect parameters and the dilaton potential \eqref{eq:minstrindilpot}. The advantage of this choice is that the quantum theory has a good semiclassical limit. 

It would be nice to understand the origin of this mismatch in \eqref{eq:minstrindilpot} from a detailed evaluation of the dilaton-gravity path integral. In order to do this it might be important to fill in the gaps regarding how the minimal model appearing in the minimal string is equivalent to a time-like Liouville theory (for some progress in this direction see \cite{KapecMahajan}).

\subsection*{Negativities in spectral density} 

It was noticed in \cite{Maxfield:2020ale} and \cite{Witten:2020wvy} that the disk density of states for JT gravity with defects can become negative if the defect weight $\lambda$ is too large, above some angle-dependent critical value. It is an open question to understand what resolves this problem and the matrix integral formulation of the theory might help answer this. 
\begin{figure}
\begin{center}
\begin{tikzpicture}
  \draw[->] (-2, 0) -- (3, 0) node[right] {$E$};
  \draw[->] (0, -1) -- (0, 3) node[above] {$\rho(E)$};
  \draw[scale=1, domain=-1.2:1.6, samples=1000, smooth, variable=\x, blue] plot ({\x}, {((\x+1.2)^(0.5))*(\x-0.72)*(\x +0.12) });
  \draw[thick] (-1.2,-0.1) -- (-1.2,0.1);
\draw (-1.2,-0.5) node {\small $E_0$};
\end{tikzpicture}
\end{center}
\vspace{-0.5cm}
\caption{Disk density of states for a matrix model with string equation $u(u^2-1)+\lambda=x$, dual to the only deformation of the $(2,5)$ minimal string. For concreteness we used $\lambda=0.5$, giving $E_0 = -1.2$.} \label{fig:negly}
\end{figure}
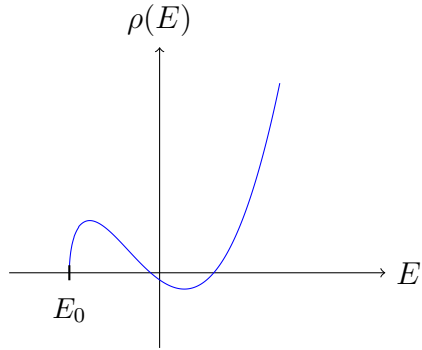 

One complication is that the matrix integral associated to JT gravity with defects is difficult and requires turning on an infinite number of operators in the matrix potential. In this section we would like to conclude with the observation that similar negativities already appears in the $(2,p)$ minimal string at finite $p$, and is thus inherited by the general class of dilaton-gravity theories derived from the $p \to \infty$ limit.

We will present the simplest case we found, which is the $(2,5)$ minimal string, corresponding to the Lee-Yang CFT coupled to 2D gravity. Besides the identity this CFT contains only one non-trivial operator $\mathcal{O}_{1,2}$, and this gives rise to the only possible deformation. We study the minimal string with the following deformation term in the action $\delta I \sim \lambda \int \mathcal{O}_{1,2} \hspace{0.1cm} e^{2 a_2 \phi}$. After some rescalings, the string equation associated to this theory is $\mathcal{F}(u) = u(u^2-1) + \lambda$. From this string equation we can compute the disk density of states using \eqref{eq:rhotoF} and we find that it becomes negative at some energies for negative enough $\lambda$. See figure \ref{fig:negly} for an example. This model corresponds to a simple matrix potential which can be easily studied to see whether there is a phase transition to a two-cut model as suggested in \cite{Witten:2020wvy} (another approach to this issue has been taken in \cite{Johnson:2020})\footnote{The negativity seems to appear for values of $\lambda$ where the interpretation of the matrix integral as a sum over random surfaces of finite size fails \cite{Kazakov:1989bc}. We thank V. Kazakov for pointing this out.}. In the case of JT gravity, understanding this regime might be necessary to study theories with a dS$_2$ region inside the bulk \cite{Anninos:2017hhn,Anninos:2018svg}.

\paragraph{Acknowledgements} 

We thank T. Budd, C. Johnson, V. Kazakov, J. Kruthoff, H. Maxfield, T. Mertens, F. Rosso and D. Stanford for useful discussions and comments on the draft. GJT would like to thank the organizers and participants of the workshop ``Matrix Models \& String Theory," August $17-21$, 2020. GJT is supported by a Fundamental Physics Fellowship. MU is supported in part by the NSF Graduate Research Fellowship Program under Grant No. DGE 1752814; by the Berkeley Center for Theoretical Physics; by the DOE, Office of Science, Office of High Energy Physics under QuantISED Award DE-SC0019380 and under contract DE-AC02-05CH11231; and by the NSF under grant PHY1820912. WWW is supported in part by the Air Force Office of Scientific Research under award number FA9550-19-1-0360.
 
\appendix

\section{Useful Formulas}\label{appformu}
In this Appendix we will collect useful identities that will be relevant in the main text. 

\subsection*{Legendre polynomials}
The Legendre polynomials $P_n(x)$ play a crucial role in the minimal string theory string equation. $P_n(x)$ is a polynomial of order $n$ defined by 
\beq
P_n(x) \equiv \frac{1}{2^n n!} \frac{d^n}{dx^n} \left( x^2-1 \right)^n,
\eeq
where $n$ is an integer. The normalization is chosen such that $P_n(1)=1$. These polynomials can also be written as a particular case of hypergeometric function 
\beq
P_{n}(x)= {}_{2}F_1\left(-n,n+1,1,\frac{1-x}{2}\right).
\eeq
The following relation is also useful
\beq
\frac{P_{n+1}(x)-P_{n-1}(x)}{2n+1} = (x-1) \hspace{.7mm} {}_{2}F_1\left(-n,n+1,2,\frac{1-x}{2}\right) 
\eeq
and the derivative of this combination of polynomials is given by
\beq
\frac{d}{dx} \left( P_{n+1}(x)-P_{n-1}(x) \right) =  (2n+1) P_{n}(x).
\eeq
This is precisely the combinations appearing in the string equation of the undeformed minimal string.

\paragraph{Integrals:} The following integrals are particularly relevant since they appear in the calculation of the minimal string density of states from the string equation. Having in mind this application, we call $p=2m-1$ and introduce an arbitrary parameter $\kappa$. Then the following identities hold 
\beq
\int_\kappa^E \frac{du}{\sqrt{E-u}} P_{m-1}\left(\frac{u}{\kappa}\right) = \frac{2\sqrt{2\kappa}}{p} \sinh\left( \frac{p}{2}{\rm arccosh} \left(\frac{E}{\kappa}\right)\right),
\eeq
and, taking $L$ to be an integer, 
\beq\label{appeqidintdp}
\int_\kappa^E \frac{du}{\sqrt{E-u}} \partial_u P_{L}\left(\frac{u}{\kappa}\right) = \frac{\sqrt{2\kappa}}{\sqrt{E^2-\kappa^2}}\cosh\left(\frac{2L+1}{2} {\rm arccosh}\left(\frac{E}{\kappa}\right) \right)-\frac{1}{\sqrt{E-\kappa}}.
\eeq

When computing disk partition functions the two integral representation of Bessel functions are very useful 
\beq
 \int_1^\infty dt \, e^{-s t} \,\frac{2}{p} \sinh \left( \frac{p}{2}~ {\rm arccosh} (t) \right) = \frac{1}{s} K_{p/2}(s)
\eeq
and 
\beq\label{idK2}
 \int_1^\infty dt \, e^{-s t} \, \frac{\cosh \left( \frac{p}{2}~ {\rm arccosh} (t) \right)}{\sqrt{t^2-1}} = K_{p/2}(s)
\eeq
Finally another useful identity 
\beq
\frac{P_m(x)-P_{m-2}(x)}{2m-1}+ \sum_{n=1}^{m-1} \frac{\lambda^n}{n!} \partial_x^{n-1} P_{m-n-1}(x) = (1-2\lambda)^{m/2}\frac{P_m(\frac{x}{\sqrt{1-2\lambda}})-P_{m-2}(\frac{x}{\sqrt{1-2\lambda}})}{2m-1}.
\eeq
This identity can be proven using the representation of the Legendre polynomial as a contour integral used in \cite{Belavin:2008kv}. 

\paragraph{Large order limit:} It will be relevant in order to take the JT limit to consider the large order limit of these Legendre polynomials. In particular we will use the following relation 
\beq
 P_{m-n-1}\Big( 1+\frac{8\pi^2}{(2m-1)^2}u \Big) \to I_0 (2\pi \alpha \sqrt{u}),~~~~n=\frac{2m-1}{2}(1-\alpha)
\eeq
where we take $m\to\infty$ with $0<\alpha<1$ and $u$ fixed. To prove this we can rewrite the Legendre polynomial as a hypergeometric function, take the $m\to\infty$ limit of the Taylor expansion in $u$, and then recognize precisely the Taylor coefficient of the Bessel function on the right hand side. 

A related useful limit will be 
\beq
\frac{2m-1}{8\pi}\left[P_{m}\Big( 1+\frac{8\pi^2}{(2m-1)^2}u \Big)-P_{m-2}\Big( 1+\frac{8\pi^2}{(2m-1)^2}u \Big) \right]\to \sqrt{u}\hspace{0.1cm}I_{1} \left( 2\pi \sqrt{u} \right),
\eeq
where we take $m\to\infty$. For a proof of this limit see \cite{Mertens:2020hbs}.

\subsection*{Bessel function identities}
The modified Bessel functions of the first kind $I_n(x)$ appear from taking the JT limit of the minimal string theory string equation. They satisfy the symmetry property $I_{-n}(x) = I_n(x)$ and the following recurrence relation
\be
\left(\frac{1}{x}\frac{d}{dx}\right)^m \Big(x^{-n} I_n(x)\Big) = x^{-(n+m)} I_{n+m}(x)
\ee
which occur in the derivation of the JT with defects string equation.

\paragraph{Integrals:} The following integral of the modified Bessel functions appears in the calculation of the JT with defects density of states from the string equation
\be \label{struveIntegral}
\int_0^E \frac{du}{2\sqrt{E-u}} \left( \frac{a}{\sqrt{u}} \right)^{n+1} I_{n+1}\Big(a\sqrt{u} \Big) = \sqrt{\frac{\pi}{2}} \left(\frac{a}{\sqrt{E}}\right)^{n+1/2} L_{n + 1/2}\Big(a \sqrt{E} \Big).
\ee
where $L_\nu$ is the modified Struve function, evaluated at half-integer order. These functions are related to the more familiar modified Bessel functions by
\be \label{finite_sum}
    L_{n+1/2}(x) = I_{-n-1/2}(x)-\frac{1}{2^{n}} \sqrt{\frac{2}{\pi}}\, \sum_{m=0}^{n} \frac{(-1)^m (2m)!}{n!(n-m)!} x^{n-2m-1/2}.
\ee
The Bessel functions at half-integer order actually have elementary form \cite{Moore:1991ir}
\be
    I_{-n-1/2}(x) = \frac{1}{\sqrt{2\pi x}} \sum_{m=0}^n \frac{(n+m)!}{m!(n-m)!} (2 x)^{-m} \bigg((-1)^m e^x + (-1)^{n} e^{-x}\bigg).
\ee
A useful way of writing the above summation is as follows
\be \label{eq:besselhalfinteger}
    I_{-n-1/2}(x) = (-1)^n \sqrt{\frac{2}{\pi x}} \bigg(p_n\left(1/x\right) \cosh{x} + q_{n-1}\left(1/x\right) \sinh{x}\bigg)
\ee
where we define the finite polynomials
\beq \label{eq:besselhalfintegerpoly}
    p_n\left(x\right) = \sum_{\substack{0 \leq m \leq n : \\m = n \text{ mod } 2}} \frac{(n+m)!}{m!(n-m)!} (x/2)^{m} ,~~~q_n\left(x\right) = \sum_{\substack{0 \leq m \leq n+1 : \\m = n+1 \text{ mod } 2}} \frac{(n+1+m)!}{m!(n+1-m)!} (x/2)^{m}.
\eeq
It is easy to see that $p_n$ and $q_n$ have degree $n$, are vanishing for $n<0$ and have definite parity.
%\bea
%    &\text{deg}\,p_n = \text{deg} \, q_n = n, \quad & p_n = q_n = 0,\, n < 0 \\ \nonumber
%    &p_n(-x) = (-1)^n p_n(x), \quad & q_n(-x) = (-1)^n q_n(x).
%\eea

\paragraph{Laplace transforms:} The Bessel functions are related to elementary functions via the following useful Laplace transforms. When computing the partition function from the JT with defects density of states we have the following integral: for any $\nu > -1$,
\be \label{eq:besselLaplace}
\int_0^\infty dE \, e^{-\beta E} \left(\frac{a}{\sqrt{E}}\right)^{\nu} I_{\nu}\Big(a\sqrt{E}\Big) = 2^{-\nu}\beta^{-\nu-1} e^\frac{a^2}{4\beta}.
\ee
On the other hand, for any $\nu > -1/2$,
\be \label{eq:besselLaplace3}
\int_0^\infty d\varphi \, e^{-\varphi y} \left(\frac{\varphi}{\sqrt{u}}\right)^\nu I_{\nu}(\sqrt{u}\varphi) = \frac{2^\nu \Gamma(\nu+1/2)}{\sqrt{\pi}} \left(y^2-u\right)^{-\nu-1/2}.
\ee
The case $\nu = -1$ is special and evaluates to
\be \label{eq:besselLaplace2}
\int_0^\infty d\varphi\, e^{-\varphi y}\, \frac{\sqrt{u}}{\varphi} I_{-1}(\sqrt{u}\varphi) = y - \sqrt{y^2 - u}.
\ee
These expressions are used in simplifying the JT with defects string equation.

\section{Disk one-point function: Normalization} \label{appendix:onept}
In this Appendix we compare the minimal string bulk one point function calculation to the matrix model prediction originally computed by \cite{Belavin:2010ba}, finding agreement as expected. We leave out the technical details, referring the interested reader to \cite{Mertens:2020hbs,Belavin:2010ba,Seiberg:2003nm}. We mostly follow the conventions of \cite{Mertens:2020hbs} for the continuum calculations.

\paragraph{Liouville approach:} First we will compute the continuum partition function to leading order in the deformation to the action $I_{\mstr}\to I_{\mstr} - \tau_n \mathcal{T}_n$. Then we will call $Z_0^{\mstr}(\ell)$ the disk partition function in the undeformed minimal string and $Z_1^{\mstr}(\ell) = \tau_n \langle \mathcal{T}_n \rangle_\ell$ the linear order correction in $\tau_n$. In both cases the continuum calculation factorizes into a Liouville part and a minimal model matter part.

The undeformed minimal string disk partition function with boundary length $\ell$, following the conventions of \cite{Mertens:2020hbs}, is given by
\bea
Z_0^{\mstr}(\ell) &=& \frac{8 \pi}{b} \left( \pi \mu \gamma(b^2) \right)^{\frac{1}{2 b^2}} \frac{(1-b^2)}{\Gamma(b^{-2})} \int_\kappa^\infty d \mu_B~ e^{-\ell \hspace{.4mm} \mu_B} \sinh \left( \frac{1}{b^2} {\rm arccosh}\frac{\mu_B}{\kappa}  \right)\\
&=& \frac{8 \pi}{b} \left( \pi \mu \gamma(b^2) \right)^{\frac{1}{2 b^2}} \frac{1}{\Gamma(\frac{1}{b^2}-1)} \frac{1}{ \ell} K_{ \frac{1}{b^2}}(\kappa \ell). 
\eea
The linear order contribution is given by the fixed length tachyon one point function. Following the conventions of \cite{Mertens:2020hbs} it is given by 
\begin{equation}
Z_{1}^{\mstr}(\ell)= \tau_n \langle \mathcal{T}_{n} \rangle_{\ell}= \tau_n \frac{4 \pi}{b}\left(\pi \mu \gamma(b^2)\right)^{-i P /b} \frac{\Gamma\left(1+ 2 i P b\right)}{\Gamma\left( -2 i P/b\right)} K_{ \frac{2 i P}{b}}(\kappa \ell) \times \langle \mathcal{O}_{1,n} \rangle_{1,1},
\end{equation}
where $P = i (1-n b^2)/(2b)$, $K_\nu$ is the modified Bessel function of the second kind, and the matter one point function $\mathcal{O}_{1,n}$ is evaluated with identity brane boundary conditions and the identity operator normalized as $\langle \mathcal{O}_{1,1} \rangle_{1,1} = 1$. The matter one point function can be found from the modular S-matrix of the minimal model and is equal to \cite{Cardy:1989ir} 
\be
\langle \mathcal{O}_{1,n} \rangle_{1,1} = \sqrt{\frac{S^{1,n}_{1,1}}{S^{1,1}_{1,1}}} = i^{n-1} \sqrt{ \frac{\sin \left(\pi n b^2 \right)} {\sin (\pi b^2) }}.
\ee
It is convenient to write the final answer in terms of the leg-factor 
\begin{equation}\label{eq:legpole1ptfses}
\operatorname{Leg}(n)=\frac{i^{n-1}}{2} \sqrt{ \frac{\pi \gamma(n b^2)}{ \mu (\pi \mu \gamma(b^2))^n}} ~ \frac{\Gamma(\frac{1}{b^2} - 1)}{\Gamma(\frac{1}{b^2}-n)} .
\end{equation}
The normalized tachyon one point function is now given by
\be \label{eq:cont1ptfct}
\frac{Z_1^{\mstr}(\ell)}{Z_0^{\mstr}(\ell)} =\tau_n \operatorname{Leg}(n)  \frac{\kappa \ell K_{\frac{p}{2}-n}(\kappa \ell)}{K_{ \frac{p}{2}}(\kappa \ell) }, 
\ee
where we used that $b^2=2/p$ and rewrote the Liouville momentum $P$ in terms of the minimal model operator $n$. This is the continuum prediction of the normalized disk one-point function. Now we will compare it with the matrix model approach.

\paragraph{Matrix Model:} From the matrix model we will compute the correction to the disk partition function to linear order in $\lambda_n$. We expand both the density of states $\rho(E)=\rho_0(E) + \rho_1(E)+\mathcal{O}(\lambda_n^2)$ and the partition function $Z^{\mm}(\ell)=Z_0^{\mm}(\ell) + Z_1^{\mm}(\ell)+\mathcal{O}(\lambda_n^2)$. To this order the string equations for any deformation $n$ is given by   
\beq
\mathcal{F}(u)=\frac{p}{16\pi^2}\left[P_m\left(\frac{u}{\kappa}\right)-P_{m-2}\left(\frac{u}{\kappa}\right)\right] + \lambda_n P_{m-n-1}\left(\frac{u}{\kappa}\right) + \mathcal{O}(\lambda_n^2)
\eeq
The undeformed disk density of states is 
\beq\label{eq:appstringeqlinear}
\rho_0(E)= \frac{p}{2\pi \kappa} \int_{\kappa}^E \frac{du}{\sqrt{E-u}} P_{m-1}\left(\frac{u}{\kappa}\right) = \frac{2\sqrt{2\kappa}}{2\pi p}\frac{p}{16 \pi^2}\frac{p}{\kappa} \sinh \left( \frac{p}{2} {\rm arccosh} \left(\frac{E}{\kappa}\right)\right),
\eeq
and the partition function is
\beq
Z_0^{\mm}(\ell) = \int_\kappa^\infty dE\, \rho_0(E) e^{-\ell E} =\frac{\sqrt{2\kappa}}{2\pi } \frac{p^2}{16\pi^2} \frac{1}{\kappa \ell} K_{p/2}(\kappa \ell) .
\eeq
To order $\sim \lambda_n$ we get a contribution from the linear order correction to the string equation and also from the linear order correction to $E_0$ applied to the density of states coming from the zeroth order string equation. 

Lets consider first the latter. The contribution from the order $\lambda_n^0$ term in the string equation is 
\beq
 \frac{1}{2\pi } \int_{E_0}^E \frac{du}{\sqrt{E-u}} \partial_u \mathcal{F}_0(u) = \rho_0 (E) -  \frac{1}{2\pi } \int_{\kappa}^{E_0} \frac{du}{\sqrt{E-u}} \partial_u \mathcal{F}_0(u).
\eeq
Now define $E_0=\kappa + \delta E_0$, where $\delta E_0$ is of order $\lambda_n$, and expand to linear order in $\delta E_0$. Therefore the second integral above is over a small range. We can approximate   
\beq\label{qowoe}
 \frac{1}{2\pi } \int_{E_0}^E \frac{du}{\sqrt{E-u}} \partial_u \mathcal{F}_0(u) = \rho_0 (E) -  \frac{\delta E_0 \partial_u \mathcal{F}_0(\kappa)}{2\pi \sqrt{E-\kappa}} + \mathcal{O}(\lambda_n^2).
\eeq
Finally, if we expand the equation for $E_0$, the string equation $\mathcal{F}_0(E_0) +  \lambda_n \mathcal{F}_1(E_0)+\mathcal{O}(\lambda^2)=0$, to linear order in $\lambda_n$ and use the fact that $\mathcal{F}_0(\kappa)=0$, we obtain
\beq
\mathcal{F}_0(\kappa) + \delta E_0 \partial_u \mathcal{F}_0(\kappa) + \lambda_n \mathcal{F}_1(\kappa)+\mathcal{O}(\lambda_n^2)=0 ~~~\Rightarrow~~~ \delta E_0  = -\frac{ \lambda_n \mathcal{F}_1(\kappa)}{\partial_u \mathcal{F}_0(\kappa)}.
\eeq
In our case, the string equation is given by \eqref{eq:appstringeqlinear}. Therefore $\mathcal{F}_1(u) = P_{m-n-1}(u/\kappa)$ and $\mathcal{F}_1(\kappa)=1$, giving $\delta E_0 \partial_u \mathcal{F}_0(\kappa) = - \lambda_n$. Inserting this in equation \eqref{qowoe} we get the linear order correction from the shift in the edge of the spectrum as
\beq
 \frac{1}{2\pi } \int_{E_0}^E \frac{du}{\sqrt{E-u}} \partial_u \mathcal{F}_0(u) = \rho_0 (E) + \frac{\lambda_n}{2\pi \sqrt{E-\kappa}} + \mathcal{O}(\lambda_n^2).
\eeq

Adding all terms, the final answer for the linear order density of states is 
\bea
\rho_1(E)&=& \frac{\lambda_n}{2\pi} \int_{\kappa}^E \frac{du}{\sqrt{E-u}} \partial_u P_{m-n-1}\left(\frac{u}{\kappa}\right)+\frac{\lambda_n}{2\pi\sqrt{E-\kappa}} \\
&=&\lambda_n \frac{\sqrt{2\kappa}}{2\pi \sqrt{E^2-\kappa^2}} \cosh \left( \frac{p-2n}{2} {\rm arccosh} \left(\frac{E}{\kappa}\right)\right).
\ea
Notice that after a change of variables $E(s)=\kappa \cosh \left( 2 \pi b s\right)$, this equation becomes precisely the density of states \eqref{swqweqwe} obtained from the Liouville one-point function. The partition function can be easily computed using the identity \eqref{idK2}, obtaining 
\beq
Z_1^{\mm}(\ell) = \int_\kappa^\infty dE~ \rho_1(E) e^{-\ell E} = \frac{\sqrt{2\kappa}}{2\pi}\lambda_n K_{\frac{p}{2}-n}(\kappa \ell).
\eeq
Putting the two results together we can compute the ratio between the zeroth and linear order partition function
\beq\label{eq:mm1ptfct}
\frac{Z_1^{\mm}(\ell)}{Z_0^{\mm}(\ell)} = \lambda_n \frac{16 \pi^2}{p^2} \frac{\kappa \ell K_{\frac{p}{2}-n}(\kappa \ell)}{K_{\frac{p}{2}}(\kappa \ell)}.
\eeq
This is the prediction from the matrix model. 

\paragraph{Comparison:} Now we can compare the continuum result \eqref{eq:cont1ptfct} with the matrix model calculation \eqref{eq:mm1ptfct}. Demanding that both results agree gives the following identification between the parameter in the string equation $\lambda_n$ and the parameter in the action $\tau_n$ given by 
\beq\label{eq:maptaulambdaapp}
\lambda_n = \tau_n \hspace{0.1cm}{\rm Leg}(n) \frac{p^2}{16 \pi^2},
\eeq
with ${\rm Leg}(n)$ given by \eqref{eq:legpole1ptfses}. This dictionary between the string equation and the coupling of the deformation in the action can also be obtained by comparing the matrix model sphere correlation functions with the one computed by Liouville. This is done for example in \cite{Goulian:1990qr} and also \cite{Belavin:2008kv} and it matches with the identification \eqref{eq:maptaulambdaapp} obtained from the fixed length disk correlator.

\section{String equation simplification}\label{app:eqLTSE} 
In this Appendix, we will show that the string equation for JT gravity with general defects (\ref{JTdefgeneralsemul}) can be written as in (\ref{eq:invlapse})
\be
    \mathcal{F}(u) = \left. \int_\mathcal{C} \frac{dy}{2\pi i} \, e^{\varphi y} \left(y-\sqrt{y^2-u-2 W(y)}\right) \right \vert_{\varphi = 2\pi}, ~~~ W(y)= \sum_i \lambda_i e^{-2\pi(1-\alpha_i) y},
\ee
which is an inverse Laplace transform evaluated at $\varphi = 2\pi$. The contour $\mathcal{C}$ is taken to be along to the imaginary direction with all singularities to the left. For simplicity, we will restrict to a single defect species $\alpha$ and evaluate the formula by Taylor expanding in the coupling $\lambda$
\bea
 y-\sqrt{y^2-u-2W(y)} &=& y - \sqrt{y^2-u} + \sum_{L=1}^\infty \frac{\lambda^L}{L!} \frac{2^{L-1} \Gamma(L-1/2)}{\sqrt{\pi}} \frac{e^{-2\pi L (1-\alpha) y}}{(y^2-u)^{L-1/2}}.
\eea
Using \eqref{eq:besselLaplace2} the first term evaluates to the string equation for pure JT gravity
\be
    \left. \int_\mathcal{C} \frac{dy}{2\pi i}\, e^{\varphi y} \left(y-\sqrt{y^2-u}\right) \right \vert_{\varphi = 2\pi} = \frac{\sqrt{u}}{2\pi} I_1(2\pi \sqrt{u}).
\ee
For $L \geq 1$, we can make use of the formula \eqref{eq:besselLaplace3} to write
\be
\left(\frac{2\sqrt{u}}{\varphi}\right)^{L-1} \int_\mathcal{C} \frac{dy}{2\pi i} \, e^{\varphi y} \, \frac{2^{L-1} \Gamma(L-1/2)}{\sqrt{\pi}} (y^2-u)^{-L+1/2} = \left(\frac{\varphi}{\sqrt{u}}\right)^{L-1} I_{L-1}(\sqrt{u}\varphi)
\ee
and that for $a\geq0$
\be
f(\varphi - a)\Theta(\varphi - a) = \int_\mathcal{C} \frac{dy}{2\pi i} \,e^{\varphi y} \tilde{f}(y) e^{-a y}.
\ee
The answer at $\mathcal{O}(\lambda^L)$ is then
\be
    \left(\frac{\varphi-2\pi L(1-\alpha)}{\sqrt{u}}\right)^{L-1} I_{L-1}\Big((\varphi-2\pi L (1-\alpha))\sqrt{u}\Big)\, \Theta(\varphi - 2\pi L(1-\alpha)),~~~L\geq0.
\end{equation}
Substituting this back into the sum and setting $\varphi = 2\pi$, we find
\begin{equation}
    \mathcal{F}(u) = \frac{\sqrt{u}}{2\pi} I_1(2\pi \sqrt{u}) + \sum_{L=1}^{\lfloor \frac{1}{1-\alpha} \rfloor} \frac{\lambda^L}{L!} \left(\frac{2\pi (1-L(1-\alpha)}{\sqrt{u}}\right)^{L-1} I_{L-1}\Big(2\pi(1 - L (1-\alpha))\sqrt{u}\Big)
\end{equation}
which is precisely the string equation for a single species (\ref{eq:onegdef}). The string equation for an arbitrary number of species follows from an analogous calculation, resulting in the replacements $\lambda^L \to \Pi_i \lambda_i^{\ell_i}$ and $L(1-\alpha) \to \sum_i \ell_i(1-\alpha_i)$ with $\ell_i = 0,1,\dots,L$ and $L = \sum_i \ell_i$. This is precisely (\ref{JTdefgeneralsemul}), as claimed.

Although we performed a Taylor expansion in $\lambda$, we note that the truncation of the series due to the step functions implies that this derivation holds for arbitrary large values of the couplings.

\section{Density of states with general defects}\label{app:pertexpexamples}
In this Appendix, we will explicitly evaluate the disk density of states for JT gravity with general defects (\ref{JTDefDiskDOS}), which we reproduce here for convenience
\be
\rho(E) = \frac{e^{S_0}}{2\pi} \sum_{L=0}^{\lfloor \frac{1}{1-\alpha} \rfloor} \frac{\lambda^L}{L!} \int_{E_0}^E \frac{du}{2\sqrt{E-u}}  \left( \frac{2\pi(1-L(1-\alpha))}{\sqrt{u}} \right)^{L} I_{L}\Big( 2\pi \big(1-L(1-\alpha)\big)\sqrt{u} \Big),
\ee
from which we can compute the partition function
\be
Z(\beta) = \frac{1}{2\pi} \int_{E_0}^\infty dE \, \rho(E) e^{-\beta E}.
\ee
For the case of $E_0 = 0$, we will evaluate these expressions exactly to all orders in $\lambda$. For $E_0 \neq 0$, we will only write down the perturbative expansion.

\subsection*{Case $E_0 = 0$}
Let us first consider the case with $E_0 = 0$. A single defect species can never satisfy this condition, so $E_0 \neq 0$ always in that case. Nevertheless, for the sake of notational convenience, we will perform the calculations with a single defect species and make the replacements $\lambda^L \to \Pi_i \lambda_i^{\ell_i}$ and $L(1-\alpha)\to \sum_i \ell_i(1-\alpha_i)$, with $\ell_i = 0,1,\dots,L$ and $L = \sum_i \ell_i$, at the end.

In this case, the integral in the density of states formula can be evaluated exactly using (\ref{struveIntegral}) and (\ref{finite_sum}) and the result is
\bea
\rho(E) &=& \frac{e^{S_0}}{2\pi} \sum_{L=0}^{\lfloor \frac{1}{1-\alpha} \rfloor} \frac{\lambda^L}{L!} \sqrt{\frac{\pi}{2}} \left(\frac{\aL}{\sqrt{E}}\right)^{L-1/2} \Bigg[I_{-L+1/2}\Big(\aL \sqrt{E} \Big) \nonumber \\
&& \qquad -\sum_{m=0}^{L-1} \frac{(-1)^m (2m)!}{(L-1)!(L-1-m)!} \left(\aL \sqrt{E} \right)^{L-2m-3/2}\Bigg]. \label{eq:JTDefDOSZeroEdge}
\eea
This form of the density of states will be useful when we discuss the $E_0 \neq 0$ case in the perturbative coupling regime in the next section. %It is useful to write down the first few terms
%\bea
%&&e^{-S_0} \rho(E) = \frac{\sinh\left(2\pi\sqrt{E} \right)}{4\pi^2} -  \frac{\lambda}{2\pi} \left(\frac{\cosh\left(2\pi\alpha\sqrt{E} \right)}{\sqrt{E}} + \frac{1}{\sqrt{E}} \right) + \frac{\lambda^2}{2\pi} \left(- \frac{\cosh\left(2\pi\left(1-2\alpha\right)\sqrt{E} \right)}{2E^{3/2}} \right. \nonumber \\
%&& ~~~~~~~~~\left.+ \frac{\pi(1-2\alpha) \sinh\left(2\pi\left(1-2\alpha\right)\sqrt{E} \right)}{E}  + \frac{1-2\pi^2 E \left(1-2\alpha\right)^2}{2E^{3/2}} \right) \Theta(2\pi(2\alpha-1)) \nonumber \\
%&& + \frac{\lambda^3}{2\pi} \left(\frac{\left(3+4\pi^2\left(2-3\alpha\right)^2 E \right)\cosh\left(2\pi\left(2-3\alpha\right)\sqrt{E} \right)}{6E^{5/2}} -\frac{\pi(2-3\alpha) \sinh\left(2\pi\left(2-3\alpha\right)\sqrt{E} \right)}{E^2} \right. \nonumber \\
%&& ~~~~~~~~~\left. + \frac{-3+2\pi^2 \left(2-3\alpha\right)^2 E-2\pi^4\left(2-3\alpha\right)^4 E^2}{6E^{5/2}} \right) \Theta(2\pi(3\alpha-2)) + O(\lambda^4).
%\eea
From \eqref{eq:besselhalfinteger} and \eqref{eq:besselhalfintegerpoly}, we see that the low energy behavior at $\mathcal{O}(\lambda^L)$ is such that a divergent contribution of $\cosh\big(\#\sqrt{E}\big)/E^{L-1/2}$ $\sim 1/E^{L-1/2}$ in the first term is precisely cancelled by a contribution of $-1/E^{L-1/2}$ in the second term, leading to the expected square-root behavior
\be
\rho(E) = e^{S_0} \sum_{L=0}^{\lfloor \frac{1}{1-\alpha} \rfloor} \frac{\lambda^L}{(L!)^2} 2^L\pi^{2L}(1-L(1-\alpha))^{2L} \sqrt{E} + \mathcal{O}(E^{3/2}).
\ee
These expressions generalize those found in \cite{Maxfield:2020ale,Witten:2020wvy}.

The partition function can be computed from the Laplace transform of (\ref{eq:JTDefDOSZeroEdge}) given in \eqref{eq:besselLaplace} and the exact answer is
\bea
Z(\beta) &=& \frac{e^{S_0}}{4\sqrt{\pi}} \sum_{L=0}^{\lfloor \frac{1}{1-\alpha} \rfloor} \frac{\lambda^L}{L!}  \frac{2^{L}}{\beta^{3/2-L}} \left(e^\ab - \sum_{m=0}^{L-1} \frac{1}{m!}\left(\ab\right)^m \right) \\
&=& \frac{e^{S_0}}{4\sqrt{\pi}} \sum_{L=0}^{\lfloor \frac{1}{1-\alpha} \rfloor} \frac{\lambda^L}{L!}  \frac{2^{L}}{\beta^{3/2-L}} e^\ab \left(1-\frac{\Gamma\Big(L,\ab \Big)}{\Gamma(L)}\right),
\eea
where we have used the series representation of the incomplete Gamma function at integer order
\be
\Gamma(n, x) = (n-1)! \, e^{-x} \sum_{m=0}^{n-1} \frac{x^m}{m!}.
\ee
It is interesting to note that in the large temperature regime $\beta \to 0$, the exponential terms dominate and we have
\bea
e^{-S_0} Z(\beta) &\approx& \frac{1}{4\sqrt{\pi}} \sum_{L=0}^{\lfloor \frac{1}{1-\alpha} \rfloor} \frac{\lambda^L}{L!}  \frac{2^{L}}{\beta^{3/2-L}} \, e^\ab \\
&=& \frac{e^{\frac{\pi^2}{\beta}}}{4\sqrt{\pi}\beta^{3/2}} + \lambda \frac{e^{\frac{\pi^2\alpha^2}{\beta}}}{2\sqrt{\pi}\beta^{1/2}} + \lambda^2 \frac{\beta^{1/2}}{2\sqrt{\pi}}\, e^{\frac{\pi^2(2\alpha-1)^2}{\beta}} \Theta(2\pi(2\alpha-1)) + \mathcal{O}(\lambda^3).
\eea
In fact, we will show below for the case $E_0 \neq 0$ that the exponential contributions are exact in the perturbative regime at each order of the coupling where the defect merging condition is satisfied.

\subsection*{Case $E_0 \neq 0$}
Let us now move on to the general case with $E_0 \neq 0$. We will again focus on a single defect species, though the results generalize easily to any number of species.

In this case, we need to perform the integral in the density of states from $E_0$ to $E$. Since we have already found the exact answer for the integral from $0$ to $E$, all that remains is to compute the integral from $0$ to $E_0$ and subtract it from the previous answer. In practice, this is difficult as the integral in general does not have a closed form. In addition, one needs to solve for $E_0$, defined as the largest root to the equation $\mathcal{F}(E_0) = 0$, which in this case is a highly nonlinear equation involving sums of modified Bessel functions.

To avoid these issues, we will focus only on the perturbative coupling regime where things simplify greatly. Since we know that $E_0 = 0$ for pure JT gravity, i.e. $\lambda = 0$, the leading contribution to $E_0$ must be of leading order $\mathcal{O}(\lambda)$. The integral from $0$ to $E_0$ is thus over a small range, and we can Taylor expand the integrand and evaluate the integral perturbatively in $E_0$. To extract the contribution at $\mathcal{O}(\lambda^L)$, one needs to solve for $E_0$ perturbatively up to $\mathcal{O}(\lambda^{L-1})$ from the string equation. In general, this can be done by iteratively as the solution at a given order depends only on the solution at lower orders. While all of this can be done explicitly, it is rather tedius in practice and we will not do it here.

Fortunately, there is a quick way of extracting the answer up to $\mathcal{O}\big(\lambda^{\lfloor \frac{1}{1-\alpha} \rfloor}\big)$ by demanding \eqref{eq:JTDefDOSZeroEdge} match the exact answer \eqref{eq:JTdefAlphaEqOne} at $\alpha = 1$. In that case, we found that the density of state reduces to that of pure JT gravity with the spectral edge shifted to $E_0 = -2\lambda$. The Taylor expansion in $\lambda$ is
\be
    \rho(E) = \frac{e^{S_0}}{4\pi^2} \sinh\left(2\pi \sqrt{E + 2\lambda}\right) = \frac{e^{S_0}}{2\pi} \sum_{L=0}^\infty \frac{\lambda^L}{L!} \sqrt{\frac{\pi}{2}} \left(\frac{2\pi}{\sqrt{E}}\right)^{L-1/2} I_{-L+1/2}\left(2\pi \sqrt{E}\right).
\ee
Comparing this to (\ref{eq:JTDefDOSZeroEdge}) at $\alpha = 1$, we see that the polynomials in $1/\sqrt{E}$ must be precisely the contributions of the integral from $0$ to $E_0$. We conclude that up to $\mathcal{O}\big(\lambda^{\lfloor \frac{1}{1-\alpha} \rfloor}\big)$, the density of states is
\beq
\rho(E) \approx \frac{e^{S_0}}{2\pi} \sum_{L=0}^{\lfloor \frac{1}{1-\alpha} \rfloor} \frac{\lambda^L}{L!} \sqrt{\frac{\pi}{2}} \left(\frac{\aL}{\sqrt{E}}\right)^{L-\frac{1}{2}} I_{-L+\frac{1}{2}}\Big(\aL \sqrt{E} \Big).
\eeq
Finally, the partition function is, up to the same order,
\bea
Z(\beta) = \frac{e^{S_0}}{4\sqrt{\pi}} \sum_{L=0}^{\lfloor \frac{1}{1-\alpha} \rfloor} \frac{\lambda^L}{L!} \frac{2^{L}}{\beta^{3/2-L}} \, e^\ab + \mathcal{O}\left(\lambda^{\lfloor \frac{1}{1-\alpha} \rfloor + 1}\right).
\eea
It is easy to show using the string equation that the higher order terms take the form of finite polynomials in $\beta^{-1/2}$.

\mciteSetMidEndSepPunct{}{\ifmciteBstWouldAddEndPunct.\else\fi}{\relax}
\bibliographystyle{utphys}
{\small \bibliography{references.bib}{}}

\end{document}